\documentclass{article}
\usepackage{ijcai26}

\usepackage{times}
\usepackage{soul}
\usepackage{url}
\usepackage[hidelinks]{hyperref}
\usepackage[utf8]{inputenc}
\usepackage[small]{caption}
\usepackage{graphicx}
\usepackage{amsmath}
\usepackage{amsthm}
\usepackage{booktabs}
\usepackage{algorithm}
\usepackage{algorithmic}
\usepackage[switch]{lineno}
\usepackage{enumitem} 
\usepackage{amsmath,amssymb,amsfonts}
\usepackage{textcomp}
\usepackage{xcolor}

\usepackage{multirow}
\usepackage{colortbl}
\usepackage{svg} 
\usepackage{stfloats} 
\usepackage{booktabs}
\usepackage{multirow}
\usepackage{graphicx}
\usepackage{array}

\newcommand{\vv}{\textbar}

\newtheorem{theorem}{Theorem}

\title{Cross-Paradigm Graph Backdoor Attacks with Promptable Subgraph Triggers}

\author{
Dongyi Liu$^{1,2}$
\and
Jiangtong Li\thanks{$^\#$Corresponding Authors}$^{\#,2}$
\affiliations
$^1$The Hong Kong University of Science and Technology (Guangzhou)\\
$^2$Tongji University\\
\emails
dliu587@connect.hkust-gz.edu.cn,
jiangtongli@tongji.edu.cn,
}

\begin{document}

\maketitle

\begin{abstract}
Graph Neural Networks~(GNNs) are vulnerable to backdoor attacks, where adversaries implant malicious triggers to manipulate model predictions. 
Existing trigger generators  are often simplistic in structure and overly reliant on specific features, confining them to a single graph learning paradigm, such as graph supervised learning, graph contrastive learning, or graph prompt learning.
Such paradigm-specific designs lead to poor transferability across different learning frameworks, limiting attack success rates in general testing scenarios.
To bridge this gap, we propose $\textbf{C}$ross-$\textbf{P}$aradigm $\textbf{G}$raph $\textbf{B}$ackdoor $\textbf{A}$ttacks with Promptable Subgraph Triggers~($\textbf{CP-GBA}$), which employs Graph Prompt Learning~(GPL) to synthesize transferable subgraph triggers. Specifically, we first distill a compact yet expressive trigger set into a queryable repository, jointly optimizing for class-awareness, feature richness, and structural fidelity. Furthermore, we pioneer the theoretical exploration of GPL transferability under prompt-based objectives, ensuring robust generalization to diverse and unseen test-time paradigms.
 Extensive experiments across multiple real-world datasets and defense scenarios show that CP-GBA achieves state-of-the-art attack success rates.
 Code is available at \url{https://github.com/novdream/CP-GBA}.
\end{abstract}

\section{Introduction}
\label{sec:introduction}

\begin{figure}[t] 
    \centering 
    \includegraphics[width=0.45\textwidth]{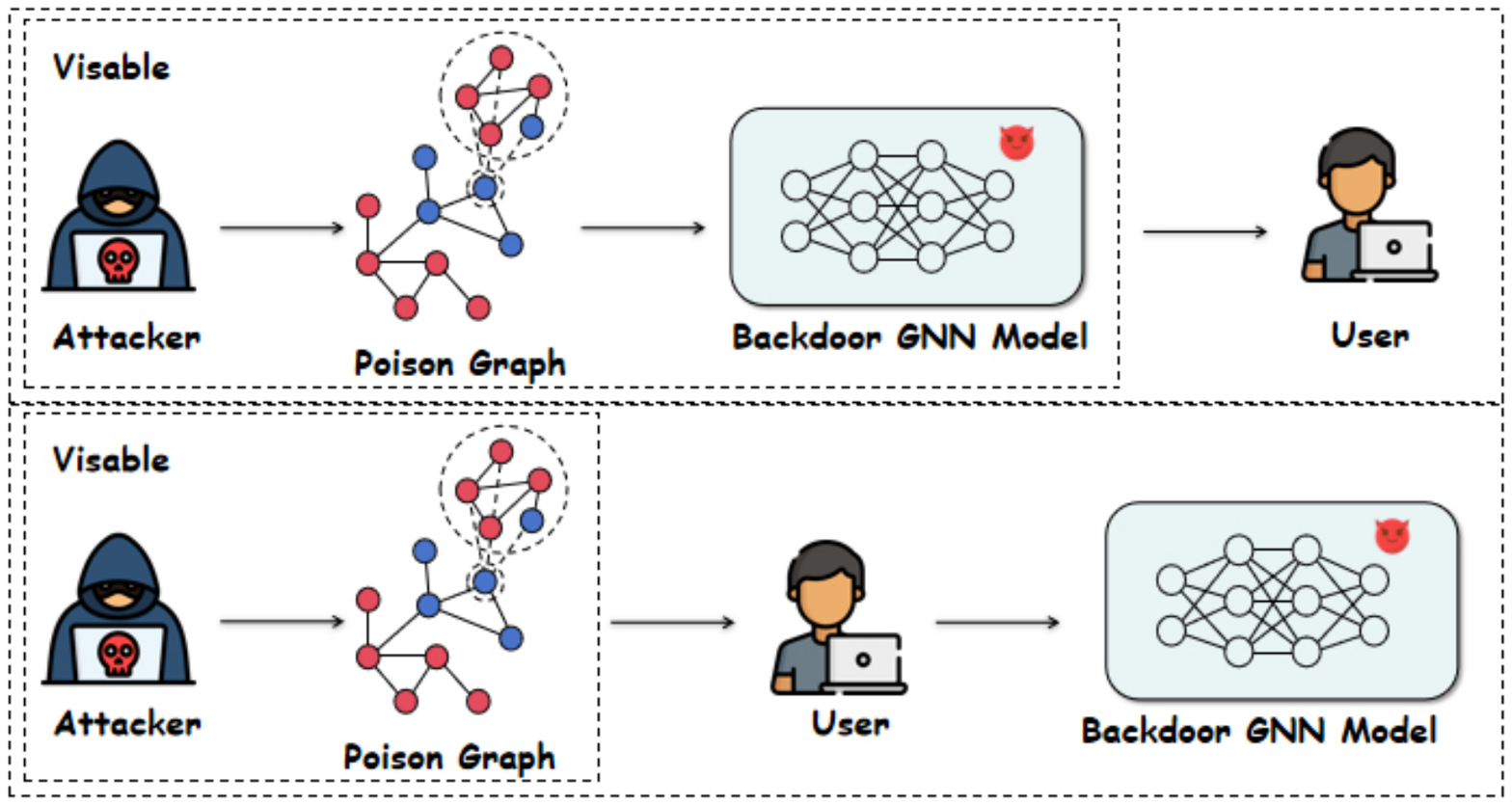} 
    \vspace{-20pt}
    \caption{\textbf{Top}: Model poisoning, where the attacker distributes a pre-trained backdoored model. \textbf{Bottom}: Data poisoning, where the victim trains a model using attacker-provided poisoned data.}
    \label{fig:scenario} 
    \vspace{-10pt}
\end{figure}

GNNs have been widely applied to analyze various types of graph-structured data in real-world applications, including social networks, molecular graphs, and financial systems~\cite{weber2019anti,fan2019graph,cheng2020spatio,bongini2021molecular}.
Their success is largely attributed to the message-passing mechanism~\cite{xu2019powerfulgraphneuralnetworks}, where nodes iteratively aggregate information from their neighbors.
This process results in node representations that preserve both local structural properties and node attributes.
Therefore, through Graph Supervised Learning~(GSL), GNNs can be applied to various supervised tasks, including node classification, link prediction, and graph classification~\cite{hamilton2017inductive,velickovic2017graph,zhang2018link,zhang2018end,jiang2019semi}.
However, GSL heavily relies on large amounts of labeled data for downstream tasks, which is often expensive and labor-intensive.
To mitigate the label dependence of traditional supervised learning, Graph Contrastive Learning~(GCL)~\cite{zhu2020deep,Wu_2024} extracts supervision from distinct augmented views, whereas Graph Prompt Learning~(GPL)~\cite{sun2022gppt,liu2023graphprompt,sun2023all,jiang2024unified} bridges pre-trained models to specific tasks via prompts.
These paradigms enable efficient training with minimal data and improve the cross-domain learning capabilities of GNNs, making them ideal for scenarios with scarce labeled data.

Although GNNs have achieved remarkable performance across various applications and diverse graph structures, prior research~\cite{zhang2023graph,dai2023unnoticeable,Zhang_2024,lyu2024cross,lin2024trojan,zhang2024robustness,li2024attackyourselfeffectiveunnoticeable} has revealed that GNNs trained under different learning paradigms are vulnerable to backdoor attacks.
For GSL, UGBA~\cite{dai2023unnoticeable} proposes a similarity loss to reduce the inconsistent similarity between triggers and attacked nodes, thereby improving trigger stealthiness.
DPGBA~\cite{Zhang_2024} further improves attack effectiveness by employing an adversarial network to address the out-of-distribution~(OOD) issue of trigger representations.
For GCL, GCBA~\cite{zhang2023graph} is designed to implant backdoors during the contrastive learning phase, showing that attacks can succeed even with limited label information.
For GPL, CrossBA~\cite{lyu2024cross} employs a hinge loss approach to maximize the similarity between the trigger and target node representations, while minimizing their similarity with clean node representations. 
This attack strategy optimizes both the triggers and GNNs, making the backdoor effective for downstream cross-task applications.
TGPA~\cite{lin2024trojan} further improves the attack success rate by optimizing triggers and task headers within a bi-level optimization framework under the assumption of a frozen GNN encoder.

\begin{table}[t]
\centering

\setlength\tabcolsep{3pt}
\resizebox{\columnwidth}{!}{%
\begin{tabular}{lcccccc}
\toprule
Paradigm & Model & Clean  & GTA & UGBA & DPGBA\\
\midrule
GSL & GCN & 85.1 & 88.8 \textbar\ 85.1 & 93.1 \textbar\ 85.1 &92.3 \textbar\ 85.0\\
GSL & GAT & 84.9 & 88.1 \textbar\ 84.9 & 92.5 \textbar\ 85.2 &92.6 \textbar\ 85.1\\
GCL & GRACE & 72.1 & 31.5 \textbar\ 62.2  & 60.9 \textbar\ 65.2 & \ 2.3 \textbar\ 67.4\\
GPL & GraphPrompt & 43.3 & 21.3 \textbar\ 37.7  & 38.2 \textbar\ 42.6 & 19.6 \textbar\ 42.5\\
\bottomrule
\end{tabular}
}
\caption{Results of existing graph backdoor attacks trained by GSL-based GCN model on Pubmed in diferent attack scenarios(Attack Success Rate (\%) \textbar\ Clean Accuracy (\%))}
\label{table:intro}
\vspace{-10pt}
\end{table}

Despite notable progress in cross-domain and cross-task backdoor capabilities, most prior works still focus on the model poisoning scenario in Fig.~\ref{fig:scenario} and attack a single learning paradigm, lacking transferability to different attack scenarios.
For example, in social network scenarios, attackers can only poison the social network data by creating virtual users~(the triggers). 
Subsequently, legitimate users crawl this public social network data to locally train backdoor models, which are designed to learn user representation embeddings for downstream tasks like recommendation systems.
However, a critical gap arises when the victim's training paradigm differs from the attacker's surrogate
Diverse learning paradigms induce distinct feature representations and distributions~\cite{xu2020understandinggraphembeddingmethods}, causing severe inconsistencies in the trigger feature space that diminish attack effectiveness.
Moreover, the trigger generators trained for graph backdoor attacks often have simple architectures and rely on node-specific features, limiting their ability to generate triggers with structure-aware diversity and feature richness.
As a result, they struggle to maintain transferability and effectiveness across diverse learning paradigms.
In Tab.~\ref{table:intro}, we validate the failure of backdoor attacks across learning paradigms on Pubmed dataset.

Therefore, in this paper, we study the novel and critical problem of designing transferable graph backdoor attacks across multiple learning paradigms in \textbf{node classification tasks}.
The challenges in developing such an attack mainly involve two key aspects:
(\textbf{i}) How to ensure trigger generalization across models trained with different learning paradigms?
(\textbf{ii}) How to design triggers that effectively capture the intrinsic structure and prior knowledge of the data?
Inspired by recent researches on GPL~\cite{sun2022gppt,sun2023all,wang2024does}, we propose \underline{C}ross-\underline{P}aradigm \underline{G}raph \underline{B}ackdoor \underline{A}ttacks with Promptable Subgraph Triggers~(\textbf{CP-GBA}), a novel approach designed to improve the transferability of graph backdoor attacks across different graph learning paradigms.
Our approach first constructs a set of condensed subgraph triggers to increase trigger diversity and maintain in-distribution structural properties, as discussed in Sec.~\ref{sec:4.1}.
To ensure transferability across learning paradigms, we further employ GPL to train triggers, utilizing its theoretical transferability detailed in Sec.~\ref{sec:4.2}.
Through these steps, CP-GBA enables effective manipulation of node classifications in graph backdoor attacks.
Extensive experiments confirm that our method consistently achieves superior attack efficacy across diverse datasets and learning paradigms, even in the presence of advanced defense mechanisms. In summary, our contributions are:
\begin{itemize}[leftmargin=10pt, topsep=0pt, partopsep=0pt]
    \item \textbf{Problem}: We address a novel backdoor attack problem: generalizing attacks across graph learning paradigms.
    \item \textbf{Method}: To the best of our knowledge, we are the first to exploit GPL for training backdoor triggers through both theoretical analysis and extensive experiments.
    \item \textbf{Results}: Extensive experiments on different real-world datasets with various defense strategies show that CP-GBA outperforms SOTA graph backdoor attack methods.
\end{itemize}

\section{Related Work}
\subsection{GNNs on Node Classification}
\noindent\textbf{GNN Architectures.}
GNNs have emerged as a powerful tool for node classification by using information propagation among nodes. Early foundational works like GCN~\cite{kipf2016semi} utilize localized spectral convolutions to aggregate neighborhood information. To capture more complex dependencies, attention-based models such as GAT~\cite{velickovic2017graph} and GraphTransformer~\cite{yun2019graph} are introduced, allowing nodes to assign learnable importance weights to neighbors. Addressing scalability in large graphs, GraphSAGE~\cite{hamilton2017inductive} enables efficient inductive learning by neighbor sampling and aggregation.

\noindent\textbf{Graph Learning Paradigms.}
Beyond architectures, diverse training paradigms have evolved to address data constraints and generalization.
For \textbf{GSL}~\cite{kipf2016semi}, the model is directly optimized through supervised loss functions using high-quality labeled data, achieving success in tasks such as node classification, link prediction, and graph classification.
For \textbf{GCL}~\cite{you2020graph}, to overcome the challenge of scarce labeled data, models aim to learn robust representations by maximizing the agreement between different augmented views of the same graph without supervision.
GraphCL~\cite{you2020graph} uses random graph augmentations to generate multiple views and learns node embeddings by contrasting positive and negative pairs via the InfoNCE loss.
CCA-SSG~\cite{zhang2021canonical} further refines the loss function using canonical correlation analysis.
For \textbf{GPL}~\cite{sun2022gppt}, to enhance cross-domain generalization under data scarcity, models adopt a ``pre-training, prompt-tuning'' strategy~\cite{yu2024hgprompt} to adapt pre-trained GNNs to downstream tasks. GPF~\cite{jiang2024unified} and GraphPrompt~\cite{liu2023graphprompt} utilize \textit{prompt-as-tokens} by appending learnable vectors to input features or latent spaces, respectively. All-in-one~\cite{sun2023all} employs \textit{prompt-as-graphs} by injecting a learnable subgraph structure where tokens connect to original nodes to unify diverse tasks.


\subsection{Backdoor Attacks on GNN}\label{sec:Backdoor}
Backdoor attacks against GNNs~\cite{yang2025graphneuralbackdoorfundamentals,ding2025spear} typically involve injecting malicious triggers into the training graph and associating them with a predetermined target label. 
As a result, when GNNs trained on the backdoored graph encounter test samples containing these triggers, they produce attacker-desired predictions.
These attacks can be categorized based on learning paradigms.
In GSL-based backdoor attacks, Zhang~\emph{et al.}~\cite{zhang2021backdoor} propose to inject universal triggers into training samples via a subgraph-based approach with limited attack success rate. 
Building on this, Xi~\emph{et al.}~\cite{xi2021graph} introduce a technique for generating adaptive triggers, customizing perturbations for individual samples to improve attack effectiveness.
Dai~\emph{et al.}~\cite{dai2023unnoticeable} propose a poisoned node algorithm to maximize the attack budget and includes an adaptive trigger generator to produce triggers with high cosine similarity to the target node.
Zhang~\emph{et al.}~\cite{Zhang_2024} further consider the OOD problem in triggers and employs a GAN loss to generate in-distribution backdoor triggers.
For GCL-based backdoor attacks, Zhang~\emph{et al.}~\cite{zhang2023graph} are the first to systematically investigate attacks across different contrastive learning stages, validating their efficacy.
For GPL-based backdoor attacks, Lyu~\emph{et al.}~\cite{lyu2024cross} propose a cross-context attack that uses a prompt-based mechanism to optimize the trigger graph and poison the pretrained GNN.
Lin~\emph{et al.}~\cite{lin2024trojan} propose a finetuning-resistant graph prompt poisoning method without poisoning the pretrained GNN.

However, existing methods focus on the intra-paradigm setting, where the surrogate model and the backdoored model belong to the same learning paradigm, resulting in attacker that are highly dependent on that specific paradigm.
In contrast, our CP-GBA is characterized by two key aspects:
(i) we tackle a new problem: transferring triggers across different paradigms under a realistic threat model, where attackers possess minimal prior knowledge and are unaware of the downstream learning paradigm.
(ii) we are the first to investigate training a set of condensed subgraph triggers with GPL, improving their generalization and transferability.

\section{Preliminary}
\subsection{Notaions}
\label{sec:notations}
We represent a graph as $\mathcal{G} = (\mathcal{V}, \textbf{A}, \textbf{X})$, where $\mathcal{V} = \{v_1, \dots, v_N\}$ is the node set, $\textbf{X} \in \mathbb{R}^{N \times d}$ is the node feature with $d$ as the feature dimension, and $N$ as the number of nodes.
The adjacency matrix $\textbf{A} \in \{0, 1\}^{N \times N}$ indicates node connectivity, where $\textbf{A}_{ij} = 1$ is an edge between nodes $v_i$ and $v_j$, and $\textbf{A}_{ij} = 0$ otherwise.
In this paper, we focus on the semi-supervised node classification in an inductive setting. 
Specifically, the graph $\mathcal{G}$ is divided into two disjoint subgraphs: the labeled graph $\mathcal{G}_L$ and the unlabeled graph $\mathcal{G}_U$, with $\mathcal{V}_L \cap \mathcal{V}_U = \emptyset$.
The labeled graph $\mathcal{G}_L$ is then split into three disjoint subsets: the labeled training graph $\mathcal{G}_T$, the validation graph $\mathcal{G}_{Va}$, and the testing graph $\mathcal{G}_{Te}$.
We denote $\mathcal{G}_{Tr} = \mathcal{G}_U \cup \mathcal{G}_T$ as the training graph for model optimization, while $\mathcal{G}_{Va}$ and $\mathcal{G}_{Te}$ are used for validation and testing.

\subsection{Threat Model}

\textbf{\textit{Attacker’s Goal.}} 
The attacker aims to poison the training graph by injecting backdoor triggers into a small set of nodes and assigning them a predefined target class label.
The GNN trained on this poisoned graph will then associate the trigger with the target class.  
As a result, it misclassifies trigger-injected nodes at test time while maintaining normal performance on clean nodes.

\noindent
\textbf{\textit{Attacker’s Knowledge and Capability.}} 
Following prior studies~\cite{Zhang_2024}, we focus on gray-box backdoor attacks targeting node classification tasks.
In a gray-box scenario, attackers have access to a small part of training data, including node attributes, graph structure, and label information, but do not know the specific architecture or parameters of the target model.
In our work, the architecture includes both the GNN structure and the learning paradigm.
Within a predefined budget, the attacker can inject triggers and assign target labels to nodes in the training graph.


\subsection{Problem Formulation}
Our preliminary analysis in Tab.~\ref{table:intro} verifies that current attack strategies are ineffective for models under different learning paradigms. 
To overcome these limitations, we propose a novel transferable graph prompt attack trained with GPL.

We first construct a condensed subgraph trigger pool~($\mathcal{T}$) with diverse structural and feature patterns.
Inspired by GPL, we adopt its mechanism to optimize the backdoor triggers, thereby improving their transferability.
We denote $\oplus$ as the process of injecting prompts $p \in \mathcal{P}$, where $\mathcal{P}$ includes both token-based and subgraph-based prompts.  
We define $a_t(\cdot)$ as the operation of selecting and attaching a trigger from the trigger pool $\mathcal{T}$.
Moreover, $f_\theta(\mathcal{G}^i)$ denotes the embedding of the local subgraph centered at node $v_i$ using the pre-trained GNN model $f_\theta$, followed by node-level classification via the classifier $f_c(\cdot)$.
Given a clean attributed graph $\mathcal{G} = (\mathcal{V}, \textbf{A}, \textbf{X})$, let $\mathcal{V}_{Tr}$ denote the set of labeled nodes used for training with clean labels.  
Let $\mathcal{V}_P \subset \mathcal{V}_U$ be the set of poisoned nodes to which triggers are attached and the target label $y_t$ is assigned.  
The optimization of backdoor triggers under the GPL setting can be formulated as:
\begin{equation}
\begin{aligned}
    & \min_{\theta_\mathcal{T}} \sum_{v_i \in \mathcal{V}_P} l(f_c(f_\theta (a_t(\mathcal{G}_P^i \oplus p, \mathcal{T}))), y_t), \\
    & s.t. \theta_c^*,p^* = \arg \min_{\theta_c, p} \sum_{v_i \in \mathcal{V}_{Tr}} l \left( f_c \left( f_\theta \left( \mathcal{G}_{Tr}^i \oplus p \right) \right), y_i \right) \\
    &\ \ + \sum_{v_i \in \mathcal{V}_P}\! l \!\left( f_c\! \left( f_\theta \!\left(\! a_t\!\left( \mathcal{G}_{P}^i \oplus p,\! \mathcal{T} \right) \right) \right), y_t \right), 
    \ \ |\mathcal{V}_P| \leq \Delta_p,
\end{aligned}
\label{eq:target}
\end{equation}
where $l(\cdot)$ is the cross-entropy loss and $\theta_\mathcal{T}$ represents the parameters of the subgraph trigger set $\mathcal{T}$.
In the constraint, the number of poisoned nodes $|\mathcal{V}_P|$ is bounded by $\Delta_p$.

\begin{figure*}[t] 
    \centering 
    \includegraphics[width=1.00\textwidth]{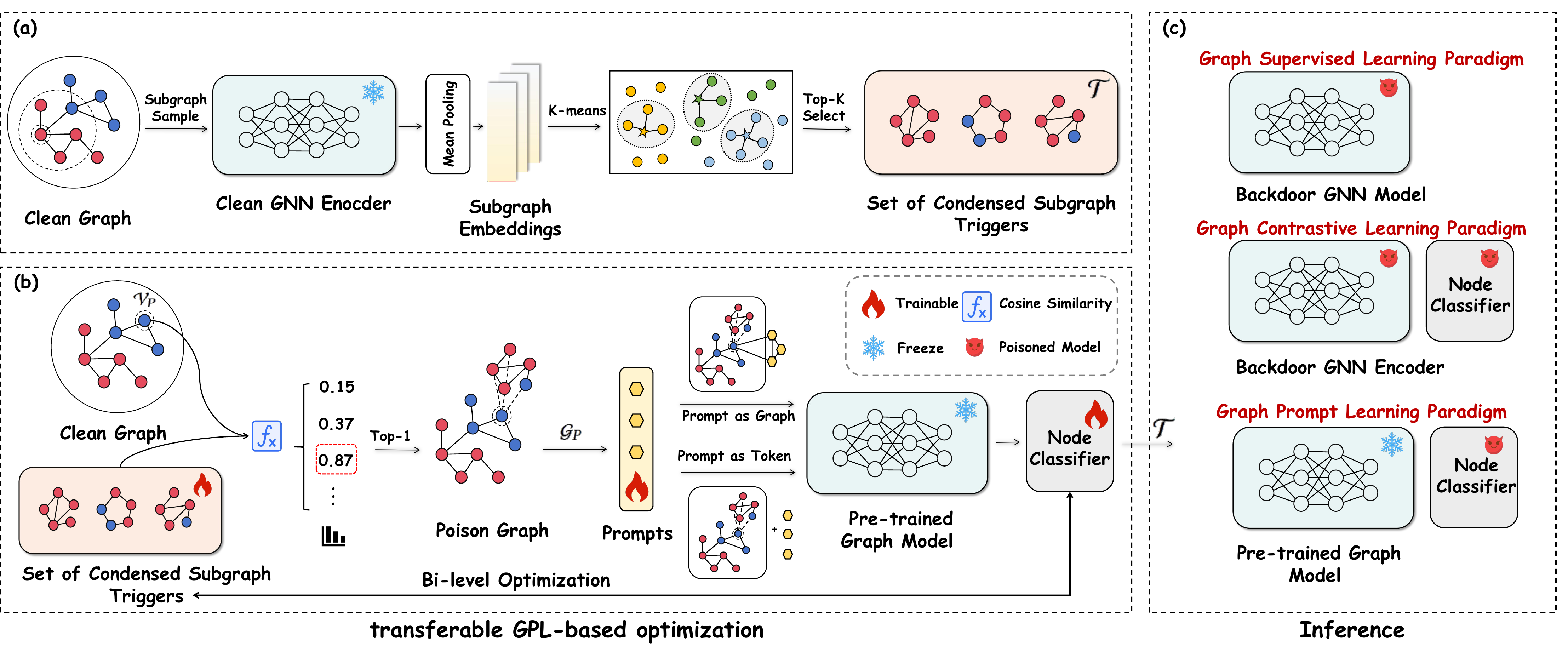} 
    \caption{The overall framework of CP-GBA, comprising three phases.
    (a) Trigger Construction: Representative subgraphs are identified via K-means clustering on the embeddings of target-class subgraphs to initialize the condensed trigger set $\mathcal{T}$.
    (b) Transferable Optimization: Triggers from $\mathcal{T}$ are injected into target nodes based on similarity, followed by joint optimization with graph prompts $\mathcal{P}$.
    (c) Inference: The transferability of the optimized backdoor triggers is evaluated across diverse learning paradigms.}
    \label{fig:main} 
\end{figure*}

\section{Methodology}
In this section, we detail our method, which optimizes Eq.~(\ref{eq:target}) to conduct transferable graph backdoor attacks, as illustrated in Fig.~\ref{fig:main}.
Our method consists of a set of condensed subgraph triggers $\mathcal{T}$, graph prompts $\mathcal{P}$ for GPL training, a frozen GNN encoder $f_\theta$, and a surrogate node classifier $f_c$.
Initially, we sample $N$ subgraphs, each containing $n$ nodes and centered on a node with the target class $y_t$, from the original graph $\mathcal{G}_{Tr}$ and select representative ones via clustering to form the initial condensed trigger set $\mathcal{T}$.
With the frozen encoder $f_\theta$ and trainable classifier $f_c$, we perform transferable GPL-based optimization to jointly train $\mathcal{T}$ and $\mathcal{P}$.  
This ensures the robust generalization of $\mathcal{T}$ across diverse GNN architectures and learning paradigms.

\subsection{The Set of Condensed Subgraph Triggers}\label{sec:4.1}
As discussed in Sec.~\ref{sec:introduction}, existing adaptive trigger generators typically adopt shallow architectures and rely heavily on node-specific features, limiting their ability to generate triggers with structural diversity and rich feature representations.
To overcome this limitation, we construct a set of condensed subgraph triggers $\mathcal{T}$ to effectively execute transferable backdoor attacks across different learning paradigms.
Initially, we train a clean two-layer GCN encoder, denoted as \( f_{\theta_{c}} \), on the training graph \( \mathcal{G}_{Tr} \) to obtain node representations, which is formulated as:
\begin{equation}
\hat{h}_i = f_{\theta_c}(\mathcal{G}_{Tr}^i) ,
\label{eq:embedding}
\end{equation}
\begin{equation}
\theta^*_c = \arg\min_{\theta_c} \sum_{v_i \in \mathcal{V}_{Tr}} l\left(\text{softmax}\left(\mathbf{W} \cdot \hat{h}_i + \mathbf{b}\right),y_i\right) ,
\label{eq:clean_encoder}
\end{equation}
where $\mathbf{W}$ denotes the trainable matrix for classification, and $\mathbf{b}$ is the bias term.  
The GCN encoder $f_{\theta_c}$ is parameterized by $\theta_c$, and $l(\cdot)$ is the cross-entropy loss.  
Here, $\hat{h}_i$ represents the embedding of node $v_i$, and $y_i$ is its ground-truth label.
The nodes of $\mathcal{V}_L$ with the target label $y_t$ are selected as central nodes.
We then employ a \textbf{B}readth-\textbf{F}irst \textbf{S}earch (BFS) algorithm to sample $N$ subgraphs, each with $n$ nodes as defined by the trigger size.
We compute representations for these $N$ sampled subgraphs using the encoder $f_{\theta_c}$ trained in Eq.~(\ref{eq:clean_encoder}).
We then apply K-means clustering to these representations to select the $K$ most representative subgraphs, which initialize $\mathcal{T} = \{t_1, t_2, \ldots, t_K\}$, where $t_i = (\textbf{X}_i^t, \textbf{A}_i^t)$ denotes the $i$-th trigger.
This approach equips $\mathcal{T}$ with category-aware, feature-rich, and structure-preserving triggers, improving both diversity and stealthiness.

\subsection{Enhancing Trigger Transferability}\label{sec:4.2}

To promote model-agnostic triggers, we not only construct $\mathcal{T}$ but also employ GPL to improve trigger generalization.
To further improve both the effectiveness and stealthiness of the attack, we adopt a bi-level optimization strategy.
In the inner loop, we optimize the surrogate classifier $f_c$ and prompts $\mathcal{P}$ with frozen GNN encoder $f_\theta$ and trigger set $\mathcal{T}$.
The outer loop freezes both $f_c$ and $\mathcal{P}$, while updating the trigger set $\mathcal{T}$.

\noindent
\textbf{Select and Inject Strategy.} 
Due to the limitations of adaptive trigger generators that rely on poisoned node features, we adopt the strategy described in Sec.~\ref{sec:4.1} to select the best-matching trigger from the condensed set $\mathcal{T}$ and inject it into the poisoned node.
Unlike UGBA, which generates sample-specific triggers, our method employs a rule-based trigger selection and injection strategy.

\begin{equation}
score_i = \frac{1}{n}\sum_{h_i \in t_i} \frac{{h}_t \cdot {h}_i}{\|{h}_t\|_2 \|{h}_i\|_2} ,
\label{eq:select}
\end{equation}
where $n$ denotes the number of nodes in a trigger, $t_i$ is the $i$-th trigger in $\mathcal{T}$, and $h_t$ is the representation of the poisoned node.  
Each $h_i$ is the representation of the $i$-th node in trigger $t_i$.
The trigger with the highest similarity score is selected for backdoor injection, denoted by the selection function $a_t(\cdot)$. 

\noindent
\textbf{Inner Loop}.
Before optimization, we randomly select a subset of unlabeled nodes from $\mathcal{V}_U$, attach triggers with the target label $y_t$, and denote this poisoned set as $\mathcal{V}_P$.
Under the empirical risk minimization setting, we fix the trigger set $\mathcal{T}$ and the encoder $f_\theta$, and train the graph prompts $\mathcal{P}$ along with the surrogate classifier $f_c$ by minimizing the loss in Eq.~(\ref{eq:surrogate_train}) on the poisoned graph.

\begin{equation}
\begin{split}
\min_{p, \theta_c} \mathcal{L}_p( \theta_c,p,\theta_\mathcal{T}) = \sum_{v_i \in \mathcal{V}_{Tr}} l \left( f_c \left( f_\theta \left( \mathcal{G}_{Tr}^i \oplus p \right) \right), y_i \right) \\
+ \sum_{v_i \in \mathcal{V}_P} l \left( f_c \left( f_\theta \left( a_t\left( \mathcal{G}_{P}^i \oplus p, \mathcal{T} \right) \right) \right), y_t \right) ,
\end{split}
\label{eq:surrogate_train}
\end{equation}
where $\theta_c$ and $\theta_\mathcal{T}$ denote the parameters of the surrogate classifier and the condensed subgraph trigger set $\mathcal{T}$, respectively.
$\mathcal{G}_{Tr}^i$ is the clean subgraph centered at node $v_i$ with ground-truth label $y_i$, and $y_t$ is the target label specified by the attacker.
$\mathcal{G}_P^i \oplus p$ represents the prompted graph.
The function $a_t(\cdot)$ selects the trigger from $\mathcal{T}$ based on the Eq.~(\ref{eq:select}).

\noindent
\textbf{Outer Loop}. 
The set of condensed subgraph triggers $\mathcal{T}$ is then optimized to mislead the surrogate classifier $f_c$, such that the frozen GNN encoder $f_\theta$ produces embeddings for nodes in $\mathcal{V}_U$ that are classified as the target label $y_t$ when attached with a selected trigger and prompted with $p$.
Formally, the objective is defined as:
\begin{equation}
\mathcal{L}_{Trans} = \sum_{v_i \in \mathcal{V}_U} l\left(f_c\left(f_\theta \left(a_t\left(\mathcal{G}_P^i \oplus p, \mathcal{T}\right)\right)\right), y_t\right) ,
\label{eq:outloop_1}
\end{equation}
Through Eq.~(\ref{eq:outloop_1}), the trigger set $\mathcal{T}$ inherits both category-specific knowledge and transferability similar to that of prompts.
Although $\mathcal{T}$ is sampled from the original graph and condensed via Eq.~(\ref{eq:select}), preserving structural consistency and improving stealthiness, it remains crucial to model both the connections between trigger and target nodes, and the internal connectivity among trigger nodes.
To address this, we define a stealthiness loss:
\begin{equation}
\mathcal{L}_{Ste} = \sum_{v_i \in \mathcal{V}_P} \sum_{(\textbf{x}_j, \textbf{x}_k) \in \mathcal{E}_t^i} \max\left(0, \tau_{sim} - \frac{\mathbf{x}_j \cdot \mathbf{x}_k}{\|\mathbf{x}_j\|_2 \|\mathbf{x}_k\|_2}\right) ,
\label{eq:hidden}
\end{equation}
where $\mathcal{E}_t^i$ denotes the set of edges between the injected trigger nodes and node $v_i$, $\tau_{sim}$ is a similarity threshold, and $\mathbf{x}_j$, $\mathbf{x}_k$ are the feature vectors of nodes $v_j$ and $v_k$.

The loss $\mathcal{L}_{p}$ in Eq.~(\ref{eq:surrogate_train}) optimizes the surrogate classifier $f_c$ to classify clean nodes correctly while also predicting $y_t$ for poisoned nodes.
Meanwhile, $\mathcal{L}_{\text{Trans}}$ in Eq.~(\ref{eq:outloop_1}) improves the transferability of the trigger set $\mathcal{T}$, and $\mathcal{L}_{\text{Ste}}$ in Eq.~(\ref{eq:hidden}) improves its stealthiness.
The final bi-level optimization objective can be formulated as:
\begin{align}
\min_{\theta_\mathcal{T}} \mathcal{L}_t(\theta_c^*, p^*, \theta_\mathcal{T}) = \mathcal{L}_{Trans} + \lambda \mathcal{L}_{Ste} \label{eq:final_loss}, \\
\text{s.t.} \quad \theta_c^*, p^* = \arg \min_{\theta_c,p} \mathcal{L}_p(\theta_c,p, \theta_\mathcal{T}), \nonumber
\end{align}
where $\lambda$ is a trade-off coefficient that balances transferability and stealthiness. 

\subsection{Why It Works}
In this section, we explore the transferability of CP-GBA from the theoretical aspect of GPL. 
\begin{theorem}
\label{theorem:1}
    In node-level, the model GNN $f$, which is trained with a large amount of high-quality data, has the ability to map any node in graph $\mathcal{G}_i$, known or unknown, to all feature spaces surjectively~(\emph{i.e}, $f:\mathcal{G}_i\rightarrow \mathbb{R}^d$, where d is the class number dimension.).
\end{theorem}

\begin{theorem}
\label{theorem:2}
    Let \( f_{\theta} \) be a GNN model trained on upstream datasets \( D_{\text{up}} \) with frozen parameters \( (\theta) \); let \( T_{\text{dow}} \) be the downstream task and \( C \) is an optimal function to \( T_{\text{dow}} \). Given any graph \( \mathcal{G}_{\text{ori}} \), \( C(\mathcal{G}_{\text{ori}}) \) denotes the optimal embedding vector to the downstream task (\emph{i.e.}, can be parsed to yield correct results for \( \mathcal{G}_{\text{ori}} \) in the downstream task), then there always exists a bridge graph \( G_{\text{bri}} \) such that \( f_{\theta}(\mathcal{G}_{\text{bri}}) = C(\mathcal{G}_{\text{ori}}) \).
\end{theorem}

Definition of the bridge graph:  
\begin{equation}
G_{bri}=G_{ori} \oplus \mathcal{T}
\end{equation}
Given a pre-trained model $f$, the downstream graph $\mathcal{G}_{\text{ori}}$ augmented with prompt $p$ yields the output representation $\hat{h} = f(\mathcal{G}_{\text{ori}} \oplus p)$.
Unlike prior backdoor attacks~\cite{dai2023unnoticeable,lyu2024cross}, our framework leverages the surjective mapping property of the pre-trained model $f$ (Theorem~\ref{theorem:1}), ensuring access to the continuous feature space $\mathbb{R}^d$.
Furthermore, Theorem~\ref{theorem:2} guarantees the existence of a bridge graph that elicits optimal downstream representations via prompt augmentation.
\textbf{Building on these guarantees, GPL-based optimization enables the trigger to function as a learnable approximation of this theoretical bridge graph.}
By dynamically adapting the trigger to minimize the prompt-tuning loss, our method effectively aligns the poisoned graph with the optimal bridge structure. This alignment empowers the trigger to inherit the intrinsic transferability of prompts, ensuring robust generalization across diverse downstream tasks, architectures, and learning paradigms.

\definecolor{shallow_blue}{rgb}{0.8235294117, 0.9490196078, 0.9490196078} 
\definecolor{shallow_yellow}{rgb}{1.0, 0.953, 0.792} 
\definecolor{red1}{rgb}{1.0, 0.0, 0.0}
\begin{table*}[t]
\centering

\setlength\tabcolsep{8pt}
\resizebox{\linewidth}{!}{%
    \begin{tabular}{l|c|cc|cc|cc|cc|cc}
        \toprule
        \multirow{2}{*}{\textbf{Dataset}}  & \multirow{2}{*}{\textbf{Defense}} & \multicolumn{2}{c}{\textbf{SBA}} & \multicolumn{2}{|c}{\textbf{GTA}} & \multicolumn{2}{|c}{\textbf{UGBA}} & \multicolumn{2}{|c}{\textbf{DPGBA}} & \multicolumn{2}{|c}{\textbf{CP-GBA}} 
        \\
        \cmidrule{3-12}
        & & \multicolumn{1}{c|}{ACC(AD)} & \multicolumn{1}{c|}{ASR} &  \multicolumn{1}{c|}{ACC(AD)} & \multicolumn{1}{c|}{ASR} &  \multicolumn{1}{c|}{ACC(AD)} & \multicolumn{1}{c|}{ASR}  &  \multicolumn{1}{c|}{ACC(AD)} & \multicolumn{1}{c|}{ASR}  &  \multicolumn{1}{c|}{ACC(AD)} & ASR 
        \\
        \midrule
        \multirow{4}{*}{Cora} & None & 0.57(+0.04) & 0.50 & 0.57(+0.04) & 0.50 & 0.59(+0.02) & \cellcolor{shallow_yellow}{0.63} & 0.60(+0.01) & {0.44} & 0.64(-0.03) & \cellcolor{shallow_blue}{0.96}  \\
        & Prune & 0.58(+0.03) & 0.41 & 0.57(+0.04) & 0.39 & 0.60(+0.01) & 0.47 & 0.57(+0.04) &\cellcolor{shallow_yellow}{0.48} & 0.64(-0.03) & \cellcolor{shallow_blue}{0.96}  \\
        & OD & 0.58(+0.03) & 0.39 & 0.57(+0.04) & 0.38 & 0.58(+0.03) & 0.47 & 0.58(+0.03) & \cellcolor{shallow_yellow}{0.49} & 0.64(-0.03) & \cellcolor{shallow_blue}{0.96} \\
         & RIGBD & 0.58(+0.03) & 0.17 & 0.57(+0.04) & 0.20 & 0.59(+0.02) & 0.21 & 0.58(+0.03) & \cellcolor{shallow_yellow}{0.22} & 0.64(-0.03) & \cellcolor{shallow_blue}{0.89} \\
        \midrule
        \multirow{4}{*}{Pubmed}& None & 0.46(+0.25) & 0.62 & 0.49(+0.22) & 0.68 & 0.73(+0.02) & \cellcolor{shallow_yellow}{0.70} & 0.72(+0.03) & 0.67 & 0.71(+0.04) & \cellcolor{shallow_blue}{0.96}  \\
        & Prune & 0.44(+0.27) & 0.35 & 0.45(+0.26) & 0.39 & 0.72(-0.01) & {0.63} & 0.73(-0.02) & \cellcolor{shallow_yellow}{0.68} & 0.71(+0.00) & \cellcolor{shallow_blue}{0.97}  \\
        & OD & 0.45(+0.26) & 0.39 & 0.46(+0.25) & 0.36 & 0.72(-0.01) & \cellcolor{shallow_yellow}{0.69} & 0.72(-0.01) & 0.67 & 0.71(+0.00) & \cellcolor{shallow_blue}{0.96} \\
        & RIGBD & 0.44(+0.27) & 0.22 & 0.45(+0.26) & 0.24 & 0.72(-0.01) & 0.39 & 0.72(-0.01) & \cellcolor{shallow_yellow}{0.45} & 0.71(+0.00) & \cellcolor{shallow_blue}{0.90} \\
        \midrule
        \multirow{4}{*}{Facebook} & None & 0.70(-0.02) & 0.22 & 0.67(+0.01) & 0.41 & 0.67(+0.01) & \cellcolor{shallow_yellow}{0.65} & 0.66(+0.02) & {0.48} & 0.68(+0.00) & \cellcolor{shallow_blue}{0.94}  \\
        & Prune & 0.68(+0.00) & 0.22 & 0.68(+0.00) & 0.23 & 0.68(+0.00) & \cellcolor{shallow_yellow}{0.77} & 0.68(-0.03) & {0.51} & 0.67(+0.01) & \cellcolor{shallow_blue}{0.95}  \\
        & OD & 0.68(+0.00) & 0.20 & 0.69(-0.01) & 0.37 & 0.69(-0.01) & \cellcolor{shallow_yellow}{0.73} & 0.68(+0.00) &{0.53} & 0.67(+0.01) & \cellcolor{shallow_blue}{0.95} \\
        & RIGBD & 0.67(+0.01) & 0.17 & 0.68(+0.00) & 0.26 & 0.68(+0.00) & \cellcolor{shallow_yellow}{0.47} & 0.67(+0.01) &{0.39} & 0.67(+0.01) & \cellcolor{shallow_blue}{0.88} \\
       
        \midrule
         \multirow{4}{*}{OGB-arxiv}  & None & 0.44(+0.03) & 0.19 & 0.45(+0.02) & 0.24 & 0.46(+0.01) & 0.70 & 0.48(-0.01) & \cellcolor{shallow_yellow}{0.73} & 0.48(-0.01) & \cellcolor{shallow_blue}{0.92}  \\
        & Prune & 0.46(+0.01) & 0.18 & 0.46(+0.01) & 0.20 & 0.46(+0.01) & {0.66} & 0.47(+0.00) & \cellcolor{shallow_yellow}{0.69} & 0.48(-0.01) & \cellcolor{shallow_blue}{0.91}  \\
        & OD & 0.46(+0.01) & 0.16 & 0.46(+0.01) & 0.19 & 0.46(+0.01) & 0.55 & 0.47(+0.00) & \cellcolor{shallow_yellow}{0.66} & 0.48(-0.01) & \cellcolor{shallow_blue}{0.91} \\
        & RIGBD & 0.45(+0.02) & 0.15 & 0.45(+0.02) & 0.17 & 0.47(+0.00) & 0.42 & 0.48(-0.01) & \cellcolor{shallow_yellow}{0.60} & 0.48(-0.01) & \cellcolor{shallow_blue}{0.91} \\
        \bottomrule
    \end{tabular}
    }
    \caption{Graph backdoor attack results (ACC(AD)~\textbar\ ASR) under different attack scenarios. The top two performances are highlighted.}\label{tab:main_result}
\end{table*}

\begin{table}[t]
\centering

\setlength\tabcolsep{12pt}
\resizebox{\linewidth}{!}{%
\begin{tabular}{l|cccc}
\toprule
\textbf{Datasets} & \textbf{Nodes} & \textbf{Edges} & \textbf{Features} & \textbf{Classes} \\
\midrule
Cora & 2,708 & 5,429 & 1,443 & 7 \\
Pubmed & 19,717 & 44,338 & 500 & 3 \\
Facebook & 22,470 & 342,004 & 128 & 4 \\
OGB-arxiv & 169,343 & 1,166,243 & 128& 40\\
\bottomrule
\end{tabular}}
\caption{Dataset statistics}\label{table:dataset}
\end{table}

\section{Experiments}

\subsection{Experimental Settings}

\textbf{Datasets.}
To evaluate the effectiveness of CP-GBA, we conduct experiments on four widely used real-world datasets, \emph{i.e.}, \textbf{Cora}, \textbf{Pubmed}, \textbf{Facebook} and \textbf{OGB-arxiv}~\cite{sen2008collective,hu2020open,rozemberczki2021multi}, which serve as standard benchmarks for inductive semi-supervised node classification. The dataset statistics
are summarized in Tab.~\ref{table:dataset}.
s
\noindent\textbf{Compared Methods.}
Following a similar setting to DPGBA~\cite{Zhang_2024}, we compare {CP-GBA} with four representative graph backdoor attack methods, including {\textbf{SBA}}~\cite{zhang2021backdoor}, {\textbf{GTA}}~\cite{xi2021graph}, {\textbf{UGBA}}~\cite{dai2023unnoticeable}, and {\textbf{DPGBA}}~\cite{Zhang_2024}.
To assess the stealthiness of {CP-GBA}, we employ three representative defenses, \textbf{Prune}~\cite{dai2023unnoticeable}, \textbf{OD}~\cite{Zhang_2024}, and \textbf{RIGBD}~\cite{zhang2025robustness}, which detect backdoors based on attribute similarity, distribution anomalies, and random edge dropping, respectively.
All hyperparameters are selected based on validation performance.

\noindent\textbf{Backbone GNN Models and Learning Methods.}
Following the taxonomy in~\cite{sun2023graphpromptlearningcomprehensive,ju2024graphcontrastivelearningsurvey}, we evaluate CP-GBA across three learning paradigms: \textbf{GSL}, including standard (GAT~\cite{velickovic2017graph}, GCN~\cite{kipf2016semi}, GraphSAGE~\cite{hamilton2017inductive}, GT~\cite{yun2019graph}) and robust architectures (GNNGuard~\cite{zhang2020gnnguard}, RobustGCN~\cite{zhu2019robust}); \textbf{GCL} (GRACE~\cite{zhu2020deep}, CCA-SSG~\cite{zhang2021canonical}); and \textbf{GPL} (GPF~\cite{jiang2024unified}, GraphPrompt~\cite{liu2023graphprompt}, All-in-one~\cite{sun2023all})


\noindent\textbf{Evaluation Protocol.}
Following a similar setting to DPGBA~\cite{Zhang_2024}, we randomly select 20\% of the nodes from the original dataset to serve as test nodes for evaluation.
Among these test nodes, half are designated as target nodes for evaluating attack performance.  
The remaining half are used as clean test nodes to assess the prediction accuracy of backdoored models on clean samples.
The graph containing the remaining 80\% of nodes is used as the training graph, where 20\% of the nodes are labeled.
To evaluate the backdoor attacks, we report the average attack success rate~(ASR) on the target node set, accuracy~(ACC) on clean test nodes, clean accuracy~(CA), and accuracy drop~(AD)~\cite{zhang2023graph}, where AD measures the performance degradation compared with the clean model. 
To assess the transferability and generalizability of CP-GBA, we repeat the experiments five times on each GNN architecture and learning paradigm and report the average performance. 
\subsection{Attack Performance}
We evaluate the transferability and stealthiness of CP-GBA against baselines across four datasets and three learning paradigms under different defense strategies. 
As shown in Tab.~\ref{tab:main_result}, we summarize the following key observations:
\begin{itemize}[leftmargin=10pt, topsep=0pt, partopsep=0pt]
    \item Baselines show poor performances across different learning paradigms, which is consistent with the findings in Tab.~\ref{table:intro}, indicating that existing attack strategies are overly reliant on model structures and learning paradigms.
    \item Without applying defense strategies, CP-GBA achieves the highest ASR over all baseline methods across all datasets, showing its generalizability and transferability with respect to different model structures and learning paradigms.
    When equipped with defense strategies, CP-GBA maintains high ASR, comparable to the no-defense setting.
\end{itemize}

\begin{figure}[t] 
    \centering 
    \includegraphics[width=\linewidth]{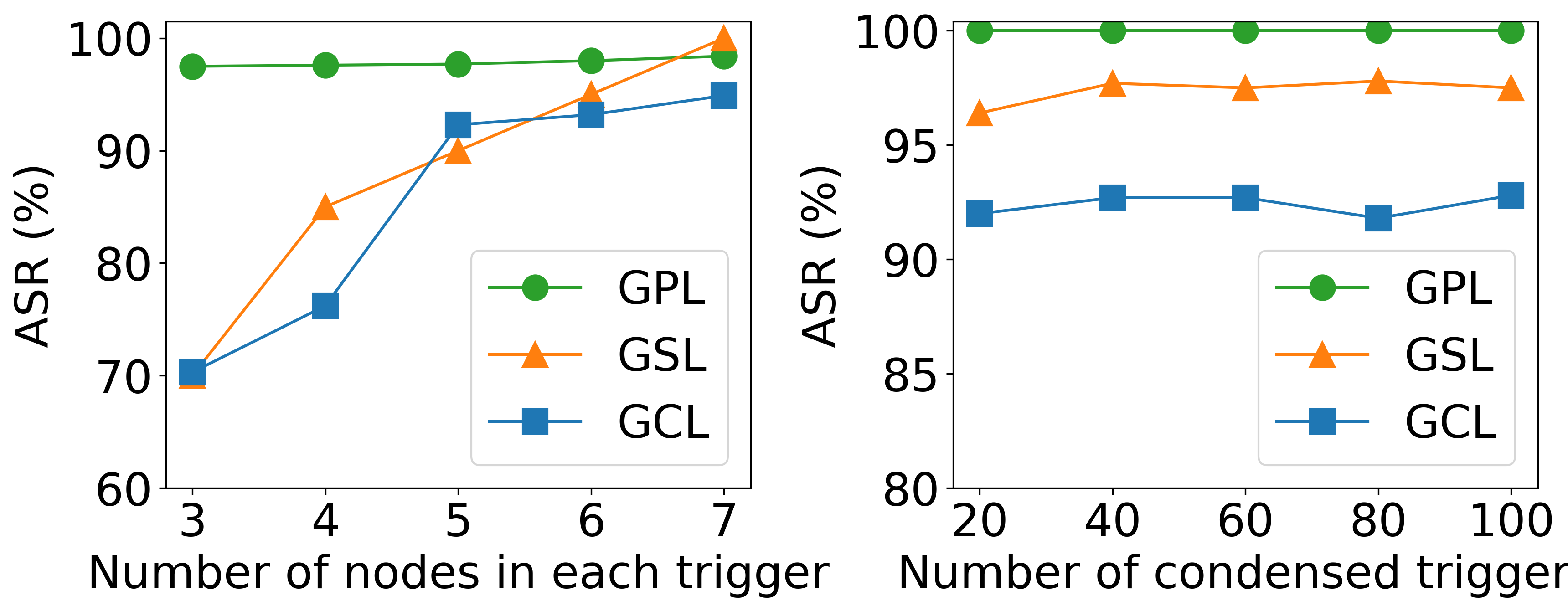} 
    \caption{ASR in different attack budgets on Cora} 
    \label{fig:size} 
    \vspace{-8pt}
\end{figure}
\vspace{-5pt}
\subsection{Impact of Attack Budget}
We investigate the impact of attack budgets by varying the trigger count in
\{20, 40, 60, 80, 100\}, and size in \{3, 4, 5, 6, 7\}, while holding all other parameters frozen.
Figure~\ref{fig:size} illustrates the ASR results on Cora, with consistent trends observed across other datasets. We report only ASR, as clean accuracy exhibits negligible fluctuations.

As the number of condensed triggers increases, the ASR shows minimal variance and remains at a high level.
This suggests that only 20 triggers are sufficient to maintain strong generalizability and effectiveness, as well as the stealthiness of the backdoor attack during graph poisoning.

As the number of nodes in each trigger increases, ASR shows a clear upward trend and plateaus when the trigger size reaches five nodes.
This indicates a 5-node trigger suffices to capture category-aware information.

\vspace{-5pt}
\subsection{Trigger Feature Distribution}
\begin{figure*}[t] 
    \centering 
    \includegraphics[width=0.95\textwidth]{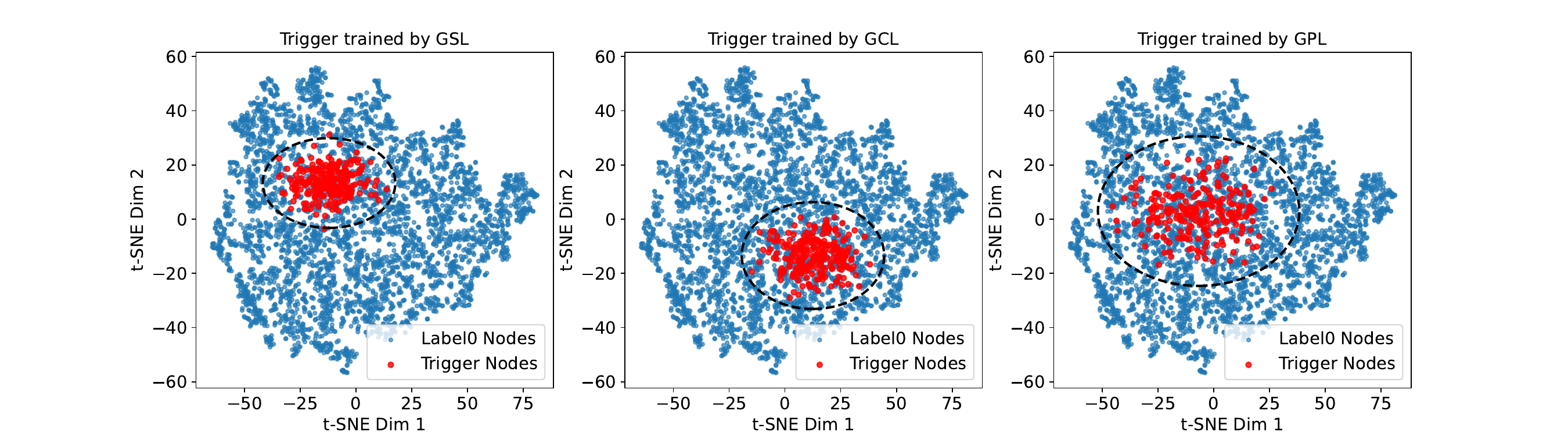} 
    \caption{t-SNE visualization of the feature embeddings for the trigger nodes and the origin nodes on the Pubmed dataset, after training with different paradigms: Graph Supervised Learning~(left), Graph Contrastive Learning~(middle), and Graph Prompt Learning~(right).}
    \label{fig:gpl_general} 
    \vspace{-8pt}
\end{figure*}

To investigate how GPL improves the generalization of triggers, we extract the embeddings of the target model for both the original nodes and the trigger nodes trained with different paradigms and visualize these embeddings using t-SNE, as shown in Fig.~\ref{fig:gpl_general}. 
We can observe that for a specific attack category, the triggers trained by GPL, while remaining close to the original node features, show a more widespread distribution that more closely aligns with the overall distribution of the original graph.
This generalization is crucial for achieving model-agnostic and paradigm-agnostic attacks, because the trigger representations avoid overfitting to a narrow region of the feature space, ensuring effectiveness across various model architectures and learning paradigms.

\begin{table*}[t]
\centering
\setlength{\tabcolsep}{10pt} 

\resizebox{0.77\linewidth}{!}{%
\begin{tabular}{l c ccc cc cc}
\toprule
\multirow{2}[2]{*}{\textbf{Data}} & \multirow{2}[2]{*}{\textbf{Defense}} & \multicolumn{3}{c}{\textbf{GPL}} & \multicolumn{2}{c}{\textbf{GCL}} & \multicolumn{2}{c}{\textbf{GSL}} \\
\cmidrule(lr){3-5} \cmidrule(lr){6-7} \cmidrule(lr){8-9}
& &  {\textbf{GPF}} &  {\textbf{G-Prompt}} &  {\textbf{AIO}} &  {\textbf{GRACE}} &  {\textbf{CCA-SSG}} &  {\textbf{GCN}} &  {\textbf{GAT}}\\
\midrule

\multirow{4}{*}{Cora}
& None   & 73.8\vv 36.2 & 94.9\vv 32.5 & 95.5\vv 34.3 & 91.5\vv 76.5 & 90.3\vv 76.9 & 83.7\vv 82.4 & 80.9\vv 83.3\\
& Prune  & 73.3\vv 36.1 & 94.1\vv 32.4 & 95.3\vv 34.1 & 90.7\vv 76.3 & 90.1\vv 76.3 & 83.9\vv 82.4 & 81.2\vv 83.8 \\
& OD     & 74.1\vv 36.4 & 93.9\vv 32.4 & 95.5\vv 33.8 & 90.9\vv 76.4 & 89.7\vv 76.9 & 84.1\vv 82.5 & 81.3\vv 83.4 \\
& RIGBD  & 71.3\vv 36.1 & 91.5\vv 32.2 & 93.9\vv 33.9 & 87.3\vv 76.4 & 87.8\vv 76.9 & 80.8\vv 82.4 & 78.6\vv 83.5 \\
\midrule
\multirow{4}{*}{Pubmed}
& None   & 93.3\vv 40.8 & 93.9\vv 44.1 & 98.6\vv 50.7 & 87.8\vv 85.2 & 87.2\vv 83.3 & 86.9\vv 84.0 & 86.4\vv 84.7\\
& Prune  & 94.0\vv 41.7 & 93.9\vv 43.7 & 98.5\vv 49.7 & 87.4\vv 85.3 & 86.6\vv 83.0 & 87.1\vv 84.2 & 87.0\vv 84.4\\
& OD     & 94.0\vv 40.9 & 94.2\vv 43.5 & 98.6\vv 49.5 & 86.9\vv 85.0 & 86.8\vv 83.2 & 87.0\vv 84.1 & 86.6\vv 84.0\\
& RIGBD  & 91.3\vv 42.7 & 90.9\vv 44.2 & 94.8\vv 50.3 & 82.1\vv 85.3 & 84.2\vv 83.3 & 83.7\vv 84.0 & 83.9\vv 84.2\\
\midrule
\multirow{4}{*}{Facebook}
& None   & 90.1\vv 50.3 & 91.8\vv 35.1 & 92.4\vv 33.3 & 85.2\vv 78.1 & 83.3\vv 80.7 & 80.5\vv 86.1 & 81.9\vv 85.7\\
& Prune  & 89.5\vv 50.3 & 92.5\vv 35.2 & 93.8\vv 33.9 & 84.5\vv 78.3 & 83.4\vv 80.6 & 79.8\vv 86.3 & 82.2\vv 85.9\\
& OD     & 90.2\vv 50.7 & 91.7\vv 35.2 & 92.4\vv 33.6 & 84.7\vv 78.3 & 83.3\vv 80.3 & 79.5\vv 86.1 & 82.0\vv 85.7\\
& RIGBD  & 87.7\vv 50.9 & 88.9\vv 35.1 & 90.3\vv 33.5 & 81.3\vv 78.3 & 80.9\vv 80.2 & 77.6\vv 86.3 & 80.1\vv 85.6\\
\midrule
\multirow{4}{*}{OGB-arxiv}
& None   & 88.6\vv 25.1 & 89.4\vv 26.1 & 91.0\vv 30.3 & 83.2\vv 53.1 & 81.4\vv 52.7 & 78.5\vv 65.1 & 77.4\vv 64.7\\
& Prune  & 88.5\vv 25.3 & 88.2\vv 26.2 & 90.8\vv 30.9 & 82.5\vv 53.3 & 81.7\vv 52.6 & 77.8\vv 65.3 & 76.9\vv 64.7\\
& OD     & 87.7\vv 25.7 & 88.7\vv 26.2 & 90.4\vv 30.2 & 82.7\vv 53.5 & 80.9\vv 52.3 & 77.5\vv 65.1 & 77.0\vv 64.7\\
& RIGBD  & 87.5\vv 25.7 & 88.8\vv 26.2 & 90.2\vv 30.7 & 81.6\vv 53.3 & 80.3\vv 52.3 & 78.1\vv 65.1 & 77.6\vv 64.7\\
\bottomrule
\end{tabular}%
} 
\caption{Backdoor attack results (ASR (\%) \textbar\ CA (\%)) on different training surrogate models from different learning paradigms.}\label{tab:three_GPL}
\end{table*}

\vspace{-5pt}
\subsection{Ablation Study}
We conduct ablation studies to investigate how GPL improves the generalizability and transferability of subgraph triggers across different model architectures and learning paradigms.
As shown in Tab.~\ref{tab:three_GPL}, we replace the subgraph trigger optimization method based on GPL with GSL and GCL.
From the experimental results, we can observe:
\begin{itemize}[leftmargin=10pt, topsep=0pt, partopsep=0pt]
    \item Under GPL optimization, All-in-one consistently outperforms token-based baselines (GPF and GraphPrompt). We attribute this superiority to the structural alignment between its subgraph prompts and our trigger design.
    Furthermore, subgraph-based prompts are likely to encode richer structural and semantic information than token-based prompts, leading to stronger transferability.
    \item The suboptimal ASR in GCL and GSL paradigms highlights their limited capacity for trigger transferability. This empirically validates the superiority of GPL in facilitating the learning of robust, transferable backdoor triggers.
\end{itemize}

\vspace{-5pt}
\section{Conclusion}
In this paper, we present both theoretical and empirical investigations on the transferability of backdoor attacks across diverse attack scenarios.
To overcome the poor transferability across attack scenarios, we identify two key challenges: (1) overreliance on the training paradigm and (2) simplistic adaptive trigger generators.
To this end, we propose CP-GBA, a transferable backdoor attack that employs a set of condensed subgraph triggers to enrich structural features and preserve distributional consistency.
Specifically, the transferability of GPL is utilized to optimize the subgraph triggers, enabling them to be model-agnostic and ensuring attack effectiveness across diverse scenarios.
Extensive experiments on real-world datasets confirm the effective performance of CP-GBA under various attack settings.

\section*{Acknowledgments}
This work is supported by the National Natural Science Foundation of China (Grant
No. 62402341).

\section*{Appendix}



\section*{Proof}
\label{sec:proof}
\begin{theorem}
\label{theorem:1}
    In node-level, the model GNN $f$, which is trained with a large amount of high-quality data, has the ability to map any node in graph $\mathcal{G}_i$, known or unknown, to all feature spaces surjectively~(\emph{i.e}, $f:\mathcal{G}_i\rightarrow \mathbb{R}^d$, where d is the class number dimension.).
\end{theorem}

\begin{proof}
Let $\Phi_{\text{GNN}}$ be a GNN with $L$ layers and injective functions. By its equivalence to the $L$-iteration WL test, this GNN maps non-isomorphic $L$-hop neighborhoods, $[G_{v,L}]$, to distinct embeddings, $h_v^{(L)}$. Thus, the map $\Phi$ is injective:
\begin{equation}
\Phi: [Gend_{v,L}] \mapsto h_v^{(L)}
\end{equation}
A local and permutation-invariant target function $f(v,G)$ depends only on the neighborhood's isomorphism class, so it can be factored as:
\begin{equation}
f(v,G) = g([G_{v,L}])
\end{equation}
Because $\Phi$ is injective, we can define a continuous function $g'$ on the GNN's output space, $\text{Im}(\Phi)$, such that:
\begin{equation}
g'(h_v^{(L)}) = g([G_{v,L}])
\end{equation}
By the Universal Approximation Theorem, there exists an MLP, $\Psi_{\text{MLP}}$, that can approximate $g'$ to arbitrary precision $\epsilon$:
\begin{equation}
\|\Psi_{\text{MLP}}(h_v^{(L)}) - g'(h_v^{(L)})\| < \epsilon \quad \text{for all } h_v^{(L)} \in \text{Im}(\Phi).
\end{equation}
The composite model $F(v,G)$ is defined as:
\begin{equation}
F(v,G) = \Psi_{\text{MLP}}(\Phi_{\text{GNN}}(v,G))
\end{equation}
This model therefore approximates $f(v,G)$, since:
\begin{equation}
\|F(v,G) - f(v,G)\| = \|\Psi_{\text{MLP}}(h_v^{(L)}) - g'(h_v^{(L)})\| < \epsilon.
\end{equation}
\end{proof}
\textbf{Corollary.}
Any surjective function $f$ mapping $d$ distinct local graph structures to $d$ distinct classes satisfies the theorem's preconditions. Since a GNN can approximate this $f$, it is therefore capable of surjectivity onto the set of $d$ class labels.

\begin{theorem}
\label{theorem:2}
    Let \( f_{\theta} \) be a GNN model trained on upstream datasets \( D_{\text{up}} \) with frozen parameters \( (\theta) \); let \( T_{\text{dow}} \) be the downstream task and \( C \) is an optimal function to \( T_{\text{dow}} \). Given any graph \( \mathcal{G}_{\text{ori}} \), \( C(\mathcal{G}_{\text{ori}}) \) denotes the optimal embedding vector to the downstream task (\emph{i.e.}, can be parsed to yield correct results for \( \mathcal{G}_{\text{ori}} \) in the downstream task), then there always exists a bridge graph \( G_{\text{bri}} \) such that \( f_{\theta}(\mathcal{G}_{\text{bri}}) = C(\mathcal{G}_{\text{ori}}) \).
\end{theorem}

\begin{proof}

For a given $G_{ori}$ and a downstream task $T_{dow}$, the embedding vector corresponding to the downstream task is formally defined as the embedding vector produced by the optimal downstream model for $T_{dow}$, which is thus uniquely determined.

Given our previous definition for the Theorem~\ref{theorem:1}, the $F_{\hat{\theta}}$ discussed here can be a surjective mapping from the graph space $\{G\}$ to $\mathbb{R}^d$. According to the properties of surjective mappings, for this particular $C(G_{ori}) \in \mathbb{R}^d$, there must exist a special graph $\hat{G}_{bri}$ such that:
\begin{equation}
F_{\hat{\theta}}(\hat{G}_{bri}) = C(G_{ori})
\end{equation}
Definition of the bridge graph:  
\begin{equation}
G_{bri}=G_{ori} \oplus \mathcal{T}
\end{equation}

Upon examining the definition of $G_{bri}$, we find that $\hat{G}_{bri} = G_{bri}$. Theorem 2 is thereby proved. 
\end{proof}

\section*{Detailed Implementation}\label{sec:implementation}
For the \textbf{subgraph trigger optimization}, the condensed triggers are first selected based on a two-layer GCN model trained on each corresponding dataset.  
These triggers are then optimized using GraphPrompt~\cite{liu2023graphprompt}, where the backbone encoder is a frozen three-layer GCN with a sum pooling layer, pre-trained by GRACE~\cite{zhu2020deep} on the Cora dataset.  
The classifier head is a trainable two-layer MLP.
For the \textbf{graph backdoor attack test stage}:  
1) \textbf{GSL}: We use two-layer variants of each corresponding GNN architecture;
2) \textbf{GCL}: We use a two-layer GCN as the encoder and a two-layer MLP as the classifier, trained in a two-stage manner on the poisoned graph;
3) \textbf{GPL}: We use a frozen two-layer GCN pretrained on the Cora dataset as the backbone encoder and a trainable two-layer MLP classifier, with three prompt nodes injected for training on the poisoned graph.
For a fair comparison, all hyperparameters are selected based on the model performance on validation set. All models are trained on a NVIDIA A6000 GPU with 48GB of memory.

\section*{Time Complexity Analysis}\label{sec:time}

Let $h$ denote the embedding dimension, $n$ the number of nodes per trigger, $K$ the number of triggers in $\mathcal{T}$, and $|\mathcal{N}_S|$ the number of candidate subgraphs to extract.

\noindent
\textbf{The $\mathcal{T}$ Construction.}  
The cost is approximately $\mathcal{O}(nh|\mathcal{N}_S| + Kh|\mathcal{N}_S| + Kh)$, accounting for subgraph extraction and K-means clustering into $K$ groups.  
As K-means clustering is the most computationally intensive step, the overall complexity is approximated as $\mathcal{O}(Kh|\mathcal{N}_S|)$.

\noindent
\textbf{Optimization.}  
Each outer iteration in the bi-level optimization consists of updating the surrogate classifier in the inner loop and optimizing the condensed trigger set $\mathcal{T}$.
The cost of updating the surrogate model is $\mathcal{O}(Nhd|\mathcal{V}|)$, where $d$ is the average node degree, $N$ is the number of inner iterations, and $|\mathcal{V}|$ is the number of training and poisoned nodes.
For trigger optimization, computing $\mathcal{L}_{Trans}$ incurs $\mathcal{O}(hd|\mathcal{V}_U|)$, where $|\mathcal{V}_U|$ is the number of unlabeled nodes.
Optimizing $\mathcal{L}_{Ste}$ costs $\mathcal{O}(h|\mathcal{V}_P||\mathcal{V}_a|)$, where $|\mathcal{V}_P|$ and $|\mathcal{V}_a|$ are the numbers of poisoned nodes and attached nodes, respectively.
Given that $|\mathcal{V}_P| \ll |\mathcal{V}|$, $|\mathcal{V}_P||\mathcal{V}_a| \ll |\mathcal{V}|$, and $|\mathcal{V}_U| \approx |\mathcal{V}|$,  
the overall time complexity per outer iteration is $\mathcal{O}((N+1)hd|\mathcal{V}|)$, which is comparable to that of UGBA.
During the backdoor attack phase, selecting and attaching a trigger incurs $\mathcal{O}(Khn + Kh + h)$, where feature extraction dominates.  
Thus, the overall cost is $\mathcal{O}(Khn)$.
This analysis indicates that \textbf{CP-GBA} scales well to large graphs.

\begin{figure}[t] 
    \centering 
    \includegraphics[width=0.45\textwidth]{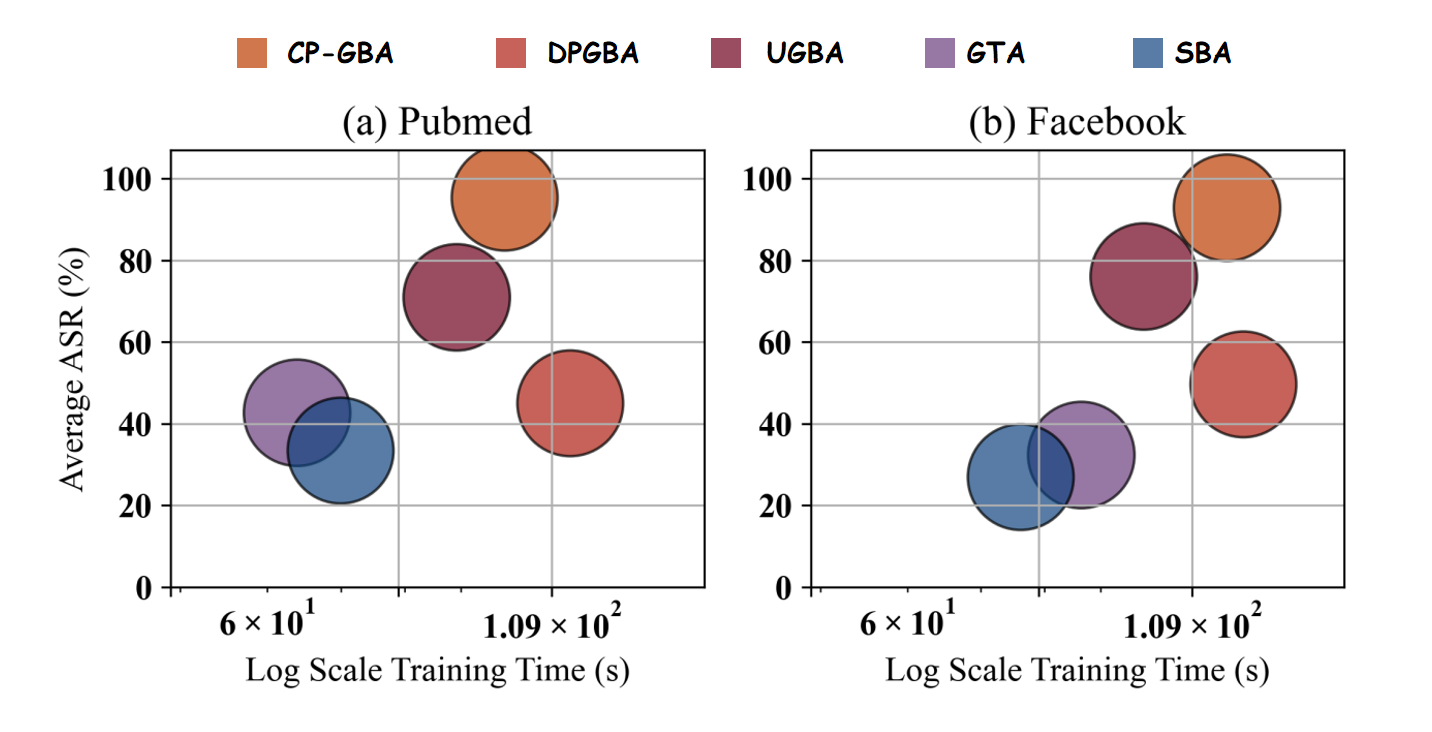} 
    \caption{Training time of triggers vs. performance}
    \label{fig:Efficiency} 
     \vspace{-6pt}
\end{figure}

We present the pseudo-code for CP-GBA in Algorithm~\ref{algorithm:1}. Furthermore, to demonstrate the practical efficiency of our optimization strategy, we report the training time of triggers across different datasets. As depicted in Fig.~\ref{fig:Efficiency}, our method achieves a favorable balance between attack performance and computational cost.

\begin{algorithm}[ht]
\caption{Algorithm of CP-GBA}
\begin{algorithmic}[1]
\REQUIRE Original Graph $\mathcal{G}$, Target Label $y_t$, Parameter $\lambda$
\ENSURE The Set of Condensed Subgraph Triggers ($\mathcal{T}$)
\STATE Initialize backdoored graph $\mathcal{G}_B = \mathcal{G}$;
\STATE Separate the training graph $\mathcal{G}_{Tr}$ from labeled graph $\mathcal{G}_L$;
\STATE Select poisoned nodes $\mathcal{V}_P$ based on the cluster algorithm from UGBA\cite{dai2023unnoticeable};
\STATE Randomly initialize node classifier $\theta_c$ and prompts $\mathcal{P}$;
\STATE Initialize $\mathcal{T}$ with $\theta_\mathcal{T}$ as parameter based on the construction of condensed riggers in Sec.4.1;
\WHILE{not converged}
    \FOR{t=1,2,...,N}
        \STATE Update $\theta_c$ by $\nabla_{\theta_c} \mathcal{L}_p$ based on Eq.~(5);
        \STATE Update $\mathcal{P}$ by $\nabla_{\mathcal{P}} \mathcal{L}_p$ based on Eq.~(5);
    \ENDFOR
    \STATE Update $\theta_\mathcal{T}$ by $\nabla_{\theta_\mathcal{T}} (\mathcal{L}_{Trans} + \lambda \mathcal{L}_{Ste})$ based on Eq.~(8);
\ENDWHILE

\RETURN $\mathcal{T}$;
\end{algorithmic}
\label{algorithm:1}
\end{algorithm}

\section*{Additional Experiments}\label{sec:Experiment}
To further assess the architectural generalizability of CP-GBA, we extend our training to include diverse surrogate models~(GraphTransformer~\cite{yun2019graph}, GAT~\cite{veličković2018graphattentionnetworks}, and GraphSAGE~\cite{hamilton2018inductiverepresentationlearninglarge}) with results detailed in Tab.~\ref{tab:more_backbone}. Additionally, we evaluate the attack's resilience against advanced defense mechanisms by testing against robust GNNs such as GNNGuard~\cite{zhang2020gnnguard} and RobustGCN~\cite{zhu2019robust} (results in Tab.~\ref{tab:more_results}). For a comprehensive overview of performance across all settings, the full results are summarized in Tab.~\ref{tab:main_result}. Experimental results on different surrogate and target models further prove the generalizability and stealthiness of our CP-GBA approach.

\begin{table}[t]
    \centering
     \scriptsize 
    \setlength{\tabcolsep}{8pt} 
    \renewcommand{\arraystretch}{1.1} 
    \caption{Average backdoor attack results (ASR(\%)\textbar\ CA (\%))where GT stands for GraphTransformer and triggers are trained by GPL with more surrogate models.}
    \begin{tabular}{lcccc}
        \toprule
        \textbf{Dataset} & \textbf{Defense} &\textbf{GT} & \textbf{GAT} & \textbf{GraphSAGE } \\
        \midrule
        \multirow{4}{*}{Cora} 
        &None & 98.1 \textbar\ 79.2 & 94.8 \textbar\ 83.3 & 98.3 \textbar\ 82.8 \\
        &Prune & 98.7 \textbar\ 78.5 & 95.2 \textbar\ 83.8 & 97.6 \textbar\ 82.9 \\
        &OD & 98.4 \textbar\ 78.6 & 94.5 \textbar\ 83.4 & 97.4 \textbar\ 83.1 \\
        &RIGBD & 94.3 \textbar\ 78.7 & 92.5 \textbar\ 83.6 & 94.7 \textbar\ 83.0 \\
         \cmidrule{1-5} 
        \multirow{4}{*}{Pubmed} 
        &None  & 96.6 \textbar\ 87.1 &94.0 \textbar\ 84.5 & 95.3 \textbar\ 85.1\\
        &Prune & 96.7 \textbar\ 87.6 & 94.1 \textbar\ 84.4 & 94.9 \textbar\ 85.0\\
        &OD    & 96.5 \textbar\ 87.2 & 93.7 \textbar\ 84.2 & 94.8 \textbar\ 84.8\\
        &RIGBD    & 93.7 \textbar\ 87.2 & 90.7 \textbar\ 84.3 & 92.2 \textbar\ 85.1\\
        \cmidrule{1-5} 
        \multirow{4}{*}{Facebook} 
        &None  & 88.6 \textbar\ 87.1 & 83.9 \textbar\ 85.2 & 84.7 \textbar\ 86.0\\
        &Prune & 88.3 \textbar\ 86.6 & 84.1 \textbar\ 85.6 & 84.0 \textbar\ 86.2\\
        &OD    & 88.5 \textbar\ 87.2 & 83.8 \textbar\ 85.5 & 84.3 \textbar\ 86.0\\
        &RIGBD    & 86.5 \textbar\ 87.2 & 82.7 \textbar\ 85.4 & 82.6 \textbar\ 86.1\\
        
        \cmidrule{1-5} 
        \multirow{4}{*}{OGB-arxiv} 
        &None  & 87.4 \textbar\ 65.5 & 87.5 \textbar\ 64.3 & 87.2 \textbar\ 65.7\\
        &Prune & 86.7 \textbar\ 65.6 & 86.6 \textbar\ 64.5 & 86.9 \textbar\ 66.2\\
        &OD    & 86.9 \textbar\ 65.2 & 86.2 \textbar\ 64.3 & 86.8 \textbar\ 66.1\\
        &RIGBD    & 84.5 \textbar\ 65.2 & 84.7 \textbar\ 64.6 & 85.8 \textbar\ 65.8\\
        \bottomrule
    \end{tabular}
    \label{tab:more_backbone}
    \vspace{-10pt}
\end{table}

\begin{table}[t]

    \centering
    \scriptsize 
\setlength{\tabcolsep}{5pt} 
\renewcommand{\arraystretch}{1.1} 
    \caption{Average backdoor attack results (ASR(\%)\textbar\ CA (\%))against more GSL-based GNNs where triggers are trained by GPL.}
    \begin{tabular}{lccccc}
        \toprule
        \textbf{Model} & \textbf{Defense} &\textbf{Cora} & \textbf{Pubmed} & \textbf{Facebook} & \textbf{OGB-arxiv}\\
        \midrule
        \multirow{4}{*}{RobustGCN} 
        &None & 99.1 \textbar\ 79.2 & 94.8 \textbar\ 84.1 & 95.2 \textbar\ 84.3 & 86.9 \textbar\ 61.3 \\
        &Prune & 99.7 \textbar\ 78.5 & 95.2 \textbar\ 83.9 & 95.4 \textbar\ 84.1 & 87.0 \textbar\ 61.1\\
        &OD & 100.0 \textbar\ 78.3 & 95.5 \textbar\ 84.2 & 95.4 \textbar\ 84.2 & 86.7 \textbar\ 61.6\\
        &RIGBD & 97.3 \textbar\ 78.8 & 92.7 \textbar\ 84.3 & 91.3 \textbar\ 84.2 & 85.4 \textbar\ 61.7\\
        \cmidrule{1-6} 
        \multirow{4}{*}{GNNGuard} 
        &None & 86.4 \textbar\ 75.7 & 81.3 \textbar\ 86.8 & 92.9 \textbar\ 85.0 & 83.7 \textbar\ 60.3\\
        &Prune & 87.1 \textbar\ 76.2 & 80.9 \textbar\ 85.9 & 94.2 \textbar\ 85.2 & 83.4 \textbar\ 60.8\\
        &OD & 87.3 \textbar\ 76.2 & 82.1 \textbar\ 86.2 & 93.0 \textbar\ 85.1 & 82.9 \textbar\ 60.7\\
        &RIGBD & 83.7 \textbar\ 76.1 & 79.8 \textbar\ 86.4 & 90.1 \textbar\ 85.3 & 81.5 \textbar\ 60.7\\
        \bottomrule
    \end{tabular}
    \label{tab:more_results}
    \vspace{-8pt}
\end{table}

\section*{Details of Compared Methods and Defense Strategies}\label{sec:details}

The compared methods are detailed as follows:
\begin{itemize}
    \item \textbf{SBA-Samp~\cite{zhang2021backdoor}}: This method targets backdoor attacks on graph classification by injecting a fixed subgraph as a trigger into the training graph for a poisoned node. The edges of each subgraph are generated using the Erdos-Renyi~(ER) model, and the node features are randomly sampled from the training graph, ensuring variability in the features of the injected subgraph.
    \item \textbf{SBA-Gen~\cite{zhang2021backdoor}}: This is a variant of SBA-Samp, SBA-Gen utilizes generated features for the trigger nodes instead of sampling from the training graph. The features for the triggers are generated from a Gaussian distribution, the parameters of which (mean and variance) are calculated based on the feature distribution of real nodes from the graph. This approach aims to create more realistic triggers for the backdoor attack.
    \item \textbf{GTA~\cite{xi2021graph}}: This method addresses backdoor attacks on both graph and node classification. It starts by randomly selecting unlabeled nodes from the clean graph as poison nodes. An adaptive trigger generator is then used to create node-specific subgraphs as triggers. The trigger generator is optimized through a bi-optimization algorithm that incorporates backdoor attack loss.
    \item \textbf{UGBA~\cite{dai2023unnoticeable}}: Similar to GTA, UGBA focuses on backdoor attacks on node classification and employs an adaptive trigger generator to generate node-specific triggers. To enhance the unnoticeability of the attack, UGBA introduces a clustering algorithm to select representative nodes as poison nodes. This method also explores the use of an unnoticeable loss function to increase the similarity between attacked nodes and generated triggers, improving the stealthiness of the backdoor attacks.
    \item \textbf{DPGBA~\cite{Zhang_2024}}: DPGBA focuses on in-domain~(ID) trigger generation for the backdoor attacks on node classification. To generate ID triggers, DPGBA introduce an out-of-distribution~(OOD) detector in conjunction with an adversarial learning strategy to generate the attributes of the triggers within distribution. This method further introduces novel modules designed to enhance trigger memorization by the victim model trained on poisoned graph.
\end{itemize}

The details of defense strategies are described as follows:
\begin{itemize}
    \item \textbf{Prune~\cite{dai2023unnoticeable}}: In this strategy, we focus on enhancing the resilience of GNNs to graph backdoor attacks by pruning edges that connect nodes with low cosine similarity. 
    This approach is based on the observation that edges created by backdoor attackers often link nodes with dissimilar features, aiming to manipulate the model's predictions subtly. 
    By pruning such edges, we can potentially disrupt the structure of the trigger inserted by the attacker, making it less effective and thus preserving the integrity of the data representation of graph. 

    \item \textbf{OD~\cite{Zhang_2024},}: Building upon the Prune strategy, this approach adds an extra defense against backdoor attacks by addressing the issue of ``dirty'' label on nodes that may have been compromised by the attacker. 
    In addition to pruning edges between dissimilar nodes, we also discard the labels of these nodes to mitigate the influence of potentially poisoned labels. 
    This dual approach helps in further safeguarding the learning process against manipulation. 
    By removing these labels, the defense mechanism reduces the risk of the model learning from and perpetuating the attacker's modifications, thereby maintaining the performance and trustworthiness of GNNs in the face of adversarial conditions.

    \item \textbf{RIGBD~\cite{zhang2025robustness}}: This method leverages the inherent robustness gap between benign and malicious graph structures. The core premise is that backdoor triggers, unlike robust benign topological features, are structurally rigid and highly sensitive to perturbations. By selectively filtering out these fragile connections during training, it effectively dismantles the trigger patterns required for the attack, thereby neutralizing the backdoor while retaining the graph's essential semantic utility.
\end{itemize}
\section*{Case Study}\label{sec:case}
\begin{figure}[t] 
    
    \centering 
    \includegraphics[width=0.45\textwidth]{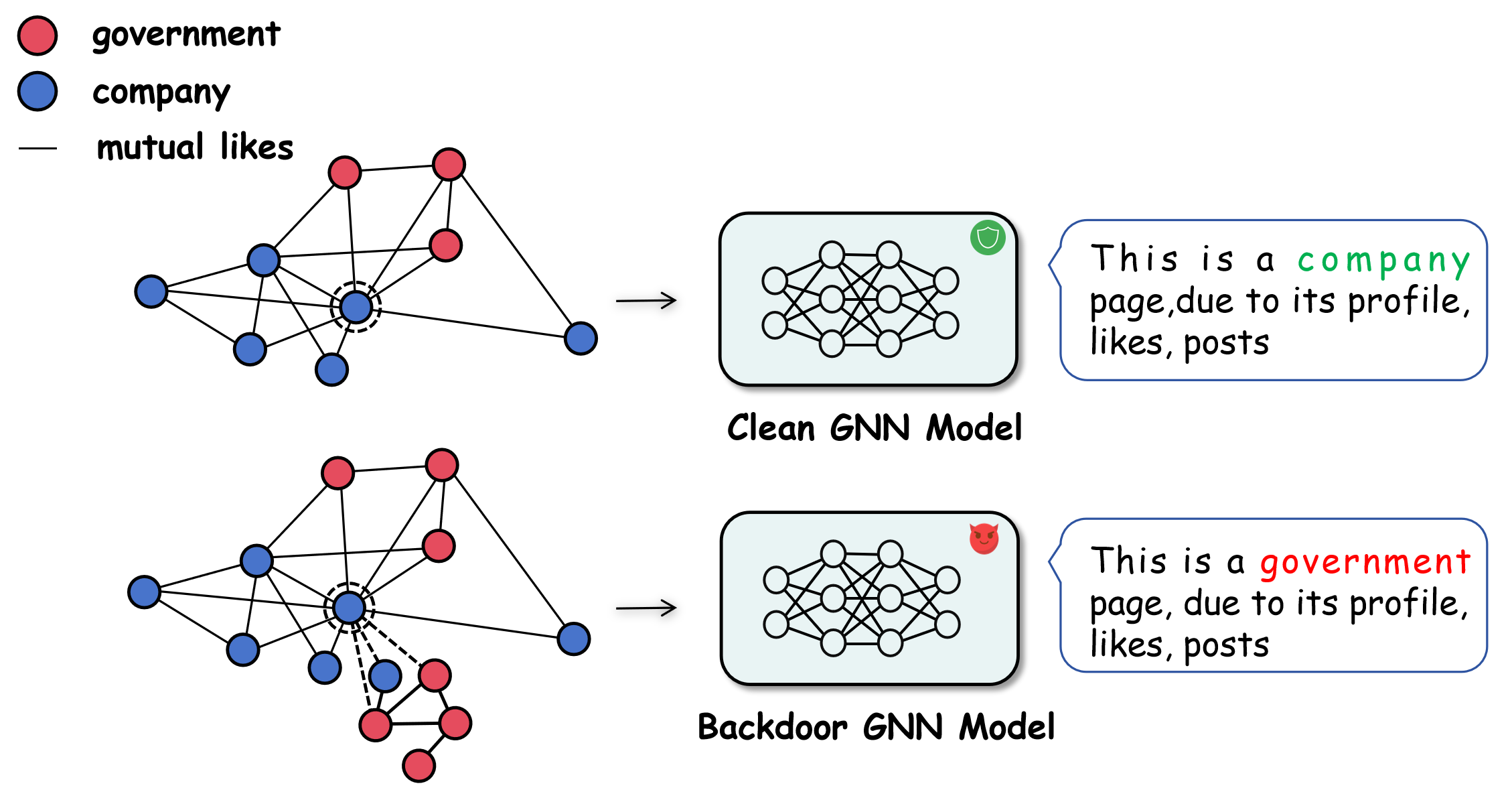} 
    \caption{Case Study on Facebook dataset}
    \label{fig:case} 
\end{figure}

In real-world scenarios such as Facebook, the set of condensed subgraph triggers may include numerous malicious user pages.
For example, a subgraph trigger may consist of 5 nodes, each representing a user page~(\emph{i.e.}, government, TV show, company, or politician).  
Edges between nodes represent mutual likes or interactions.
Attackers induce misclassification by selecting a suitable trigger from the set based on the target user's characteristics and linking it to the target.
As illustrated in Fig.~\ref{fig:case}, the blue node, originally a company page, may be mislabeled by the backdoor model as a government page after being followed by malicious accounts.

\definecolor{shallow_blue}{rgb}{0.8235294117, 0.9490196078, 0.9490196078} 
\definecolor{shallow_yellow}{rgb}{1.0, 0.953, 0.792} 
\definecolor{red1}{rgb}{1.0, 0.0, 0.0}
\begin{table*}[t!]
 \setlength\tabcolsep{2.5pt}
    \centering
     \caption{Graph backdoor attack results (ACC(AD)~\textbar\ ~ASR~) under different attack scenarios. The top two performances in ASR are highlighted in blue and yellow.} 
     %
   
    \begin{tabular}{ll|c|cc|cc|cc|cc|cc}
        \toprule
        \multirow{2}{*}{\textbf{Dataset}} & \multirow{2}{*}{\textbf{Method}} & \multirow{2}{*}{\textbf{Defense}} & \multicolumn{2}{c}{\textbf{SBA}} & \multicolumn{2}{|c}{\textbf{GTA}} & \multicolumn{2}{|c}{\textbf{UGBA}} & \multicolumn{2}{|c}{\textbf{DPGBA}} & \multicolumn{2}{|c}{\textbf{CP-GBA}} 
        \\
        \cmidrule{4-13}
        & & & \multicolumn{1}{c|}{ACC(AD)} & \multicolumn{1}{c|}{ASR} &  \multicolumn{1}{c|}{ACC(AD)} & \multicolumn{1}{c|}{ASR} &  \multicolumn{1}{c|}{ACC(AD)} & \multicolumn{1}{c|}{ASR}  &  \multicolumn{1}{c|}{ACC(AD)} & \multicolumn{1}{c|}{ASR}  &  \multicolumn{1}{c|}{ACC(AD)} & ASR 
        \\
        \midrule
        \multirow{12}{*}{Cora} & \multirow{4}{*}{GSL} & None & 0.81(+0.00) & 0.58 & 0.82(-0.01) & 0.75 & 0.82(-0.01) & 0.76 & 0.81(+0.00) & \cellcolor{shallow_yellow}{0.78} & 0.81(+0.00) & \cellcolor{shallow_blue}{0.97}  \\
        & & Prune & 0.80(+0.01) & 0.68 & 0.82(-0.01) & 0.39 & 0.82(-0.01) & 0.69 & 0.81(+0.00) &\cellcolor{shallow_yellow}{0.72} & 0.81(+0.00) & \cellcolor{shallow_blue}{0.97}  \\
        &  & OD & 0.81(+0.00) & 0.68 & 0.81(+0.00) & 0.48 & 0.82(-0.01) & 0.69 & 0.81(+0.00) & \cellcolor{shallow_yellow}{0.72} & 0.81(+0.00) & \cellcolor{shallow_blue}{0.97} \\
         &  & RIGBD & 0.81(+0.00) & 0.15 & 0.81(+0.00) & 0.21 & 0.82(-0.01) & 0.19 & 0.81(+0.00) & \cellcolor{shallow_yellow}{0.22} & 0.81(+0.00) & \cellcolor{shallow_blue}{0.89} \\
        \cmidrule{2-13}
        & \multirow{4}{*}{GCL} & None & 0.73(+0.01) & 0.30 & 0.69(+0.05) & 0.25 & 0.70(+0.04) & \cellcolor{shallow_yellow}{0.51} & 0.71(+0.03) & 0.09 & 0.76(-0.02) & \cellcolor{shallow_blue}{0.91}  \\
        & & Prune & 0.75(-0.01) & 0.31 & 0.71(+0.03) & 0.24 & 0.71(+0.03) & \cellcolor{shallow_yellow}{0.49} & 0.69(+0.05) & 0.13 & 0.76(-0.02) & \cellcolor{shallow_blue}{0.92} \\
        & & OD & 0.74(+0.00) & 0.30 & 0.71(+0.03) & 0.18 & 0.70(+0.04) & \cellcolor{shallow_yellow}{0.44} & 0.69(+0.05) & 0.07 & 0.76(-0.02) & \cellcolor{shallow_blue}{0.91}  \\
        &  & RIGBD & 0.74(+0.00) & 0.18 & 0.70(+0.04) & 0.17 & 0.70(+0.04) & \cellcolor{shallow_yellow}{0.23} & 0.70(+0.04) & 0.12 & 0.76(-0.02) & \cellcolor{shallow_blue}{0.86} \\
        \cmidrule{2-13}
        & \multirow{4}{*}{GPL} & None & 0.18(+0.12) & 0.63 & 0.21(+0.09) & 0.51 & 0.26(+0.04) & \cellcolor{shallow_yellow}{0.63} & 0.29(+0.01) & 0.46 & 0.34(-0.04) & \cellcolor{shallow_blue}{0.99} \\
        & & Prune & 0.20(+0.10) & 0.63 & 0.18(+0.12) & \cellcolor{shallow_yellow}{0.89} & 0.26(+0.04) & 0.22 & 0.22(+0.08) & 0.59 & 0.34(-0.04) & \cellcolor{shallow_blue}{0.98}  \\
        & & OD & 0.18(+0.12) & \cellcolor{shallow_yellow}{0.79} & 0.19(+0.11) & 0.47 & 0.23(+0.07) & 0.27 & 0.23(+0.07) & 0.68 & 0.34(-0.04) & \cellcolor{shallow_blue}{0.99}  \\
        &  & RIGBD & 0.19(+0.11) & 0.19 & 0.20(+0.10) & 0.21 & 0.24(+0.06) & 0.21 & 0.23(+0.07) & \cellcolor{shallow_yellow}{0.32} & 0.34(+0.04) & \cellcolor{shallow_blue}{0.91} \\
        \midrule
        \multirow{12}{*}{Pubmed} & \multirow{4}{*}{GSL} & None & 0.86(-0.02) & 0.27 & 0.87(-0.03) & 0.79 & 0.86(-0.02) & \cellcolor{shallow_yellow}{0.79} & 0.87(-0.03) & 0.66 & 0.84(+0.00) & \cellcolor{shallow_blue}{0.96}  \\
        & & Prune & 0.86(-0.02) & 0.23 & 0.87(-0.03) & 0.19 & 0.86(-0.02) & \cellcolor{shallow_yellow}{0.80} & 0.87(-0.03) & 0.68 & 0.84(+0.00) & \cellcolor{shallow_blue}{0.97}  \\
        &  & OD & 0.86(-0.02) & 0.26 & 0.86(-0.02) & 0.19 & 0.86(-0.02) & \cellcolor{shallow_yellow}{0.79} & 0.87(-0.03) & 0.66 & 0.84(+0.00) & \cellcolor{shallow_blue}{0.97} \\
        &  & RIGBD & 0.85(-0.01) & 0.17 & 0.86(-0.02) & 0.15 & 0.86(-0.02) & 0.29 & 0.87(-0.03) & \cellcolor{shallow_yellow}{0.32} & 0.84(+0.00) & \cellcolor{shallow_blue}{0.88} \\
        \cmidrule{2-13}
        & \multirow{4}{*}{GCL} & None & 0.20(+0.64) & \cellcolor{shallow_blue}{1.00} & 0.20(+0.64) & \cellcolor{shallow_blue}{1.00} & 0.84(+0.00) & 0.67 & 0.84(+0.00) & 0.23 & 0.84(+0.00) & \cellcolor{shallow_yellow}{0.93}  \\
        & & Prune & 0.20(+0.64) & \cellcolor{shallow_blue}{1.00}  & 0.20(+0.64) & \cellcolor{shallow_blue}{1.00} & 0.85(-0.01) & 0.64 & 0.84(+0.00) & 0.23 & 0.84(+0.00) & \cellcolor{shallow_yellow}{0.93} \\
        & & OD & 0.20(+0.64) & \cellcolor{shallow_blue}{1.00} & 0.20(+0.64) & \cellcolor{shallow_blue}{1.00} & 0.84(+0.00) & 0.63 & 0.83(+0.01) & 0.20 & 0.84(+0.00) & \cellcolor{shallow_yellow}{0.93}  \\
        &  & RIGBD & 0.20(+0.64) & \cellcolor{shallow_blue}{1.00} & 0.20(+0.64) & \cellcolor{shallow_blue}{1.00} & 0.83(+0.01) & 0.39 & 0.83(+0.01) & 0.19 & 0.84(+0.00) & \cellcolor{shallow_yellow}{0.86} \\
        \cmidrule{2-13}
        & \multirow{4}{*}{GPL} & None & 0.32(+0.12) & 0.58 & 0.39(+0.05) & 0.54 & 0.50(-0.06) &0.65 & 0.45(-0.01) &  \cellcolor{shallow_yellow}{0.82} & 0.44(-0.00) & \cellcolor{shallow_blue}{1.00} \\
        & & Prune & 0.27(+0.17) & 0.72 & 0.28(+0.16) & 0.58 & 0.44(+0.00) & 0.44 & 0.47(-0.03) &\cellcolor{shallow_yellow}{0.83} & 0.44(+0.00) & \cellcolor{shallow_blue}{1.00}  \\
        & & OD & 0.28(+0.16) & 0.52 & 0.31(+0.13) & 0.79 & 0.47(-0.03) & 0.65 & 0.45(-0.01) & \cellcolor{shallow_yellow}{0.85} & 0.45(-0.01) & \cellcolor{shallow_blue}{0.99}  \\
         &  & RIGBD & 0.28(+0.16) & 0.38 & 0.30(+0.14) & 0.48 & 0.47(-0.03) & 0.49 & 0.46(-0.02) & \cellcolor{shallow_yellow}{0.54} & 0.46(-0.02) & \cellcolor{shallow_blue}{0.97} \\
        \midrule
        \multirow{12}{*}{Facebook} & \multirow{4}{*}{GSL} & None & 0.88(-0.03) & 0.47 & 0.88(-0.03) & 0.68 & 0.88(-0.03) & \cellcolor{shallow_yellow}{0.80} & 0.88(-0.03) &  \cellcolor{shallow_yellow}{0.80} & 0.85(+0.00) & \cellcolor{shallow_blue}{0.92}  \\
        & & Prune & 0.87(-0.02) & 0.49 & 0.88(-0.03) & 0.13 & 0.88(-0.03) & \cellcolor{shallow_yellow}{0.80} & 0.88(-0.03) & \cellcolor{shallow_yellow}{0.80} & 0.85(+0.00) & \cellcolor{shallow_blue}{0.92}  \\
        &  & OD & 0.87(-0.02) & 0.38 & 0.88(-0.03) & 0.57 & 0.88(-0.03) & \cellcolor{shallow_yellow}{0.80} & 0.88(-0.03) & \cellcolor{shallow_yellow}{0.80} & 0.85(+0.00) & \cellcolor{shallow_blue}{0.92} \\
        &  & RIGBD & 0.86(-0.01) & 0.31 & 0.88(-0.03) & 0.38 & 0.88(-0.03) & 0.50 & 0.88(-0.03) & \cellcolor{shallow_yellow}{0.52} & 0.85(+0.00) & \cellcolor{shallow_blue}{0.85} \\
        \cmidrule{2-13}
        & \multirow{4}{*}{GCL} & None & 0.82(-0.03) & 0.18 & 0.83(00.04) & 0.23 & 0.80(-0.01) & \cellcolor{shallow_yellow}{0.84} & 0.78(+0.01) & 0.27 & 0.79(+0.00) & \cellcolor{shallow_blue}{0.92} \\
        & & Prune & 0.82(-0.03) & 0.17 & 0.83(-0.04) & 0.16 & 0.81(-0.02) & \cellcolor{shallow_blue}{0.95} & 0.78(+0.01) & 0.21 & 0.78(+0.01) & \cellcolor{shallow_yellow}{0.93} \\
        & & OD & 0.80(-0.01) & 0.19 & 0.83(-0.04) & 0.18 & 0.84(-0.05) & \cellcolor{shallow_yellow}{0.85} & 0.78(+0.01) & 0.24 & 0.79(+0.00) & \cellcolor{shallow_blue}{0.92}  \\
        &  & RIGBD & 0.80(-0.01) & 0.18 & 0.83(-0.04) & 0.18 & 0.82(-0.03) & \cellcolor{shallow_yellow}{0.63} & 0.78(+0.01) & 0.22 & 0.79(+0.00) & \cellcolor{shallow_blue}{0.87} \\
        \cmidrule{2-13}
        & \multirow{4}{*}{GPL} & None & 0.39(-0.01) & 0.01 & 0.31(+0.07) & 0.33 & 0.34(+0.04) & 0.30 & 0.33(+0.05) & \cellcolor{shallow_yellow}{0.36} & 0.39(-0.01) & \cellcolor{shallow_blue}{0.99} \\
        & & Prune & 0.35(+0.03) & 0.01 & 0.33(+0.05) & 0.41 & 0.35(+0.03) & \cellcolor{shallow_yellow}{0.56} & 0.38(+0.00) & 0.51 & 0.39(-0.01) & \cellcolor{shallow_blue}{0.99}  \\
        & & OD & 0.37(+0.01) & 0.03 & 0.35(+0.03) & 0.37 & 0.36(+0.02) & 0.53 & 0.37(+0.01) & \cellcolor{shallow_yellow}{0.56} & 0.38(+0.00) & \cellcolor{shallow_blue}{1.00}  \\
        &  & RIGBD & 0.36(+0.02) & 0.02 & 0.32(+0.06) & 0.22 & 0.34(+0.04) & 0.29 & 0.35(+0.03) & \cellcolor{shallow_yellow}{0.42} & 0.38(+0.00) & \cellcolor{shallow_blue}{0.91} \\
        \midrule
         \multirow{12}{*}{OGB-arxiv} & \multirow{4}{*}{GSL} & None & 0.58(+0.03) & 0.25 & 0.59(+0.02) & 0.29 & 0.61(+0.00) & 0.67 & 0.62(-0.01) & \cellcolor{shallow_yellow}{0.70} & 0.61(+0.00) & \cellcolor{shallow_blue}{0.87}  \\
        & & Prune & 0.59(+0.02) & 0.23 & 0.59(+0.02) & 0.24 & 0.61(+0.00) & \cellcolor{shallow_yellow}{0.60} & 0.61(+0.00) & 0.58 & 0.60(+0.01) & \cellcolor{shallow_blue}{0.87}  \\
        &  & OD & 0.59(+0.02) & 0.21 & 0.59(+0.02) & 0.21 & 0.60(+0.01) & 0.53 & 0.61(+0.00) & \cellcolor{shallow_yellow}{0.58} & 0.61(+0.00) & \cellcolor{shallow_blue}{0.86} \\
        &  & RIGBD & 0.58(+0.03) & 0.18 & 0.58(+0.03) & 0.19 & 0.61(+0.00) & 0.35 & 0.61(+0.00) & \cellcolor{shallow_yellow}{0.72} & 0.60(+0.01) & \cellcolor{shallow_blue}{0.82} \\
        \cmidrule{2-13}
        & \multirow{4}{*}{GCL} & None & 0.50(+0.03) & 0.15 & 0.50(+0.03) & 0.19 & 0.52(+0.01) & \cellcolor{shallow_yellow}{0.77} & 0.54(-0.01) & 0.75 & 0.53(+0.00) & \cellcolor{shallow_blue}{0.93}  \\
        & & Prune & 0.52(+0.01) & 0.16  & 0.50(+0.03) & 0.18 & 0.51(+0.02) & 0.72 & 0.52(+0.01) & \cellcolor{shallow_yellow}{0.76} & 0.54(-0.01) & \cellcolor{shallow_blue}{0.93} \\
        & & OD & 0.52(+0.01) & 0.14 & 0.52(+0.01) & 0.13 & 0.51(+0.02) & 0.47 & 0.53(+0.00) & \cellcolor{shallow_yellow}{0.70} & 0.53(+0.00) & \cellcolor{shallow_blue}{0.93}  \\
        &  & RIGBD & 0.52(+0.01) & 0.14 & 0.51(+0.02) & 0.14 & 0.53(+0.00) & 0.43 & 0.54(-0.01) & \cellcolor{shallow_yellow}{0.54} & 0.54(-0.01) & \cellcolor{shallow_blue}{0.97} \\
        \cmidrule{2-13}
        & \multirow{4}{*}{GPL} & None & 0.25(+0.03) & 0.18 & 0.27(+0.01) & 0.24 & 0.26(+0.02) &0.67 & 0.29(-0.01) &  \cellcolor{shallow_yellow}{0.75} & 0.31(-0.03) & \cellcolor{shallow_blue}{0.95} \\
        & & Prune & 0.27(+0.01) & 0.14 & 0.28(+0.00) & 0.18 & 0.26(+0.02) & 0.65 & 0.27(+0.01) &\cellcolor{shallow_yellow}{0.74} & 0.30(-0.02) & \cellcolor{shallow_blue}{0.94}  \\
        & & OD & 0.28(+0.00) & 0.13 & 0.27(+0.01) & 0.22 & 0.27(+0.01) & 0.65 & 0.28(+0.00) & \cellcolor{shallow_yellow}{0.70} & 0.30(-0.02) & \cellcolor{shallow_blue}{0.93}  \\
        &  & RIGBD & 0.26(+0.02) & 0.12 & 0.27(+0.01) & 0.18 & 0.28(+0.00) & 0.49 & 0.30(-0.02) & \cellcolor{shallow_yellow}{0.55} & 0.31(-0.03) & \cellcolor{shallow_blue}{0.93} \\
        \bottomrule
    \end{tabular}
   
    \label{tab:main_result}
     
\end{table*}


\section*{Discussion}\label{sec:discussion}
\subsection{Findings}

This work yields several key findings:
\begin{itemize}
    \item \textbf{Plateau for Larger Triggers:} The attack is constrained by the GNN's limited propagation range and the original graph's degree distribution. Large triggers introduce structural anomalies, compromising both efficiency and stealthiness.
    \item  \textbf{Batch Efficiency of Trigger Sets:} The trigger set achieves superior performance over MLP-generated triggers in large batches, enabled by its parallel, database-like storage and retrieval.
\end{itemize}

\subsection{Future Direction}
Building on this study, several promising directions for future research emerge:
\begin{itemize}
    \item \textbf{Cross Task Attack:} Existing graph backdoor attacks are task-specific, either node-level or graph-level predictions. Generalizing them to multi-task settings offers a promising path toward a more universal attack.
    \item \textbf{Limited Data Access:} Current backdoor attacks often require access to training data and labels. 
    Designing data-efficient attacks that operate with extremely limited or zero data access is a critical future direction, increasing their real-world threat.
\end{itemize}

\bibliographystyle{named}
\bibliography{ijcai26}

\end{document}